\documentclass{emulateapj}
\usepackage{graphicx}
\pdfoutput=1
\usepackage[usenames]{color}

        \renewcommand{\columnsep}{0.5cm}

        \renewcommand{\baselinestretch}{0.97}

\def \etal{{et al.$\,$}}
\def \eg{{\em e.g.}}
\def \gtw{\>\hbox{\lower.25em\hbox{$\buildrel >\over\sim$}}\>}
\def \ltw{\>\hbox{\lower.25em\hbox{$\buildrel <\over\sim$}}\>}
\def \cf{{\em cf.}}

\def \ie{{\em i.e.}}

\def \arcdeg{^{\circ}}
\def \cm3{cm$^{-3}$}
\def \radm2{ rad/m$^2$}
\def \dyncm2{dyn$~$cm$^{-2}$}
\def \kmps{km$~$s$^{-1}$}
\def \ergps{erg$~$s$^{-1}$}

\def \hal{H$\alpha$\,}
\def \g{\gamma}
\def \be{\begin{equation}}
\def \ee{\end{equation}}
\shortauthors{Neff, Eilek, \& Owen}
\shorttitle{Transition Regions in Cen A: A Galactic Wind}

\begin{document}

\title{The Complex North Transition Region of Centaurus A: A Galactic Wind}

\author{Susan G. Neff}
\affil{NASA's Goddard Space Flight Center, 
Laboratory for Observational Cosmology,  
Mail Code 665,  Greenbelt, Maryland, 20771}
\email{susan.g.neff@nasa.gov}
\author{Jean A. Eilek}
\affil{Physics Department, New Mexico Tech, Socorro NM 87801}
\affil{National Radio Astronomy Observatory\footnote{The 
National Radio Astronomy Observatory is a facility of 
the National Science Foundation operated under cooperative 
agreement by Associated Universities Inc.}\footnote{Adjunct 
Astronomer at the National Radio Astronomy Observatory.},
Socorro NM 87801}
\author{Frazer N. Owen}
\affil{National Radio Astronomy Observatory$^*$, Socorro NM 87801}

\baselineskip 12pt

\begin{abstract}
	
 We present deep {\it GALEX} images of NGC 5128, the parent galaxy 
of Centaurus A.  We detect a striking  ``weather ribbon'' 
of Far-UV and H$\alpha$ emission which extends more than 
35 kpc northeast of the galaxy.  This ribbon is
associated with a knotty ridge of radio/X-ray emission and is  
an extension of the previously known string of optical emission-line
filaments.  Many phenomena in the region are too short-lived to have
survived transit out from the inner galaxy;  something must be driving
them locally. We also detect Far-UV emission from the galaxy's central 
dust lane.  Combining this with previous radio and Far-IR measurements, 
we infer an
active starburst in the central galaxy which is currently forming 
stars at $\sim 2  M_{\sun}$yr$^{-1}$, and has been doing so for
50-100 Myr.    If the wind from this starburst is 
enhanced by energy and mass driven out from the AGN, 
the  powerful augmented wind can be the driver needed for
the northern weather system. 
 We argue that both the diverse weather system, 
and the enhanced radio 
 emission in the same region, result from the wind's encounter with 
 cool gas left by one of the recent merger/encounter events in 
 the history of NGC 5128.

\end{abstract}

\bigskip

\section{Introduction}
\label{Intro:begin}

The radio source Centaurus A (``Cen A'') and its parent galaxy,
NGC 5128 are the nearest active-galaxy system. At a distance
of only 3.8 Mpc, (Harris \etal 2010;  $1\arcmin \simeq 1.14 $kpc),
both the radio source Cen A and the galaxy NGC5128 can be scrutinized
with a sensitivity and resolution impossible for other active 
galaxies.

NGC 5128 is fundamentally a normal elliptical galaxy, 
dominated by an old  stellar population,
with kinematic signatures typical of other massive ellipticals
(e.g., Peng \etal 2004;  Woodley \etal 2010;  Rejkuba \etal 2011).  
However, the galaxy's optical appearance is dominated by the unusual, 
and iconic,  dust band (Dunlop 1828, Herschel 1847).
The  central dust band is the site of vigorous and ongoing star formation.  
In this paper, we show that the extended starburst
is strong enough to drive out a wind. 

On larger scales, a striking complex of optical emission-line gas, 
active star formation, cold gas and dust clouds, and overpressured
radio and X-ray knots  extends into 
the outer reaches of the galaxy, $\sim 10-35$ kpc to the NE (all
distances are given in projection on the sky).  
This system is spatially coincident with the diffuse radio emission
we described in Neff, Eilek, and Owen (2015; Paper 1).
We suggest in this paper that both structures are causally related to
an energy-carrying flow in the region.

As in Paper 1, our focus here is the inner $\sim 50$ kpc of the Cen A
/ NGC 5128 system.  We present new {\it GALEX} observations and
consider what they, together with new radio data (Paper 1), reveal
about the astrophysics of this region.  In the rest of this
section we present an introduction to 
key aspects of the system.  In Table \ref{regions} we list regions
and features discussed throughout the paper, and give approximate
distances from the galaxy's center.

\begin{table}[htb]
\caption{Important regions \& features  in Cen A \ NGC 5128}
\label{regions}
\begin{center}
\begin{tabular}{l c }
\hline
\rule{0pt}{16pt} \hspace{-8pt }
Definition &  Scale$^{\rm a}$
 \\[2pt]
  & (kpc) 
 \\[4pt]
\hline
 (North, South) Inner radio Lobes (N/S IL)  & $\sim 7$, $\sim 5.5$ 
\\[2pt]
 North Middle (radio) Lobe$^{\rm b}$ (NML) & $10-40$ 
\\[2pt]
 (North, South) Transition Regions (N/S TR) & $10-40$ 
\\[2pt]
 (North, South) Outer radio Lobes (N/S OL) & $40- 300$ 
\\[2pt]
 Weather Ribbon / Weather System   & $ 10-40$ 
\\[2pt]
Inner Filament$^{\rm c}$, Outer Filament$^{\rm c}$ & 8, 15 
\\[2pt]
\hline
\end{tabular}
\end{center}
$^{\rm a}$ Spatial scales give approximate distance from AGN, 
{\it in projection}.\\
$^{\rm b}$ We use ``North Middle Lobe'' to refer only to the radio-loud
part of the North Transition Region.\\
$^{\rm c}$ The terms ``Inner Filament'' and ``Outer Filament'' 
are well established in the
literature, and refer to specific, optically-bright emission line filaments.
\\[10pt]
\end{table}


\subsection{Large and small-scales}
\label{Intro:large_scales}

\paragraph{ The outer  radio lobes}
The radio source Cen  A was first identified with the 
galaxy NGC 5128 by the radio emission from its 
Outer Lobes (Bolton et al. 1949), which extend $\sim 600$ 
kpc end-to-end 
on the sky 
(\eg, Junkes \etal  1993;  
Feain \etal 2011).   The outer 
radio lobes are also detected as extended $\gamma$-ray 
sources (Abdo \etal 2010), thought to result from inverse
Compton scattering of Cosmic Microwave Background photons. 
Eilek (2014) argues that the dynamical age of Cen A's outer 
lobes is on the order of $\sim 1$  Gyr (assuming the lobes lie at
some finite angle to the sky plane), but that they must 
have been last energized no more than $\sim 30$ Myr ago to
keep shining in radio and $\gamma$-rays.  Any energy supplied
to the Outer Lobes by the inner galaxy must, of course, have 
moved through the middle ($\sim 50$-kpc scale) regions which we
study in this paper.

\paragraph{The central AGN}
\label{Intro:small_scales}

The massive black hole in the nucleus of NGC 5128 is 
currently active, as indicated by pc-scale radio jets, 
a northern few-kpc-scale radio and X-ray jet, and inner lobes extending 
$\sim 7$ kpc NE and $\sim 5.5$ kpc 
SW of the galactic core  (\eg, Burns \etal 1983).
Both brightness and proper motion asymmetries
of the inner jet (\eg, Tingay \etal 2001) and
Faraday depolarization of the South Inner Lobe 
(Clarke \etal 1992) suggest that the northern jet 
is approaching us while the South Inner Lobe
is aimed away from us. X-ray observations of a 
shock driven into the ISM by the Southern Inner Lobe
show the Inner Lobes are only $\sim 1-2$ Myr old 
(Croston \etal 2009, Paper 1). This suggests 
the Active Galactic Nucleus (AGN) has only recently 
restarted as a collimated outflow. 
The power currently produced by the AGN, 
several $\times 10^{43}$\ergps, is interestingly close 
to that required to power the outer lobes over
their $\sim 1$ Gyr lifetime (Eilek 2014, paper 1).

\paragraph{The central disk and inner galaxy}
\label{Intro:small_scales_SB}

The iconic dust lane in NGC 5128 is part of a thin ($\sim 250$ pc),
multiply warped gas and dust disk.
It  is a  site of active star formation.
UBV photometry of young stars (van den Bergh 1976, Dufour \etal 1979) 
and young clusters (Minitti \etal 2004),
together with H$\alpha$ and [NII] 
imaging of HII regions (Bland \etal 1987) 
and Far-Infrared (FIR) spectroscopy of the disk (Unger \etal 2000), 
indicate ongoing active star formation in the disk that has continued 
for at least 50 Myr. The disk is thought to result 
from a merger or stripping interaction, between a large elliptical galaxy 
and a smaller gas-rich system, that occurred 250$-$750 Myr ago 
(Struve \etal 2010,  Sparke \etal
1996, Quillen \etal 1993).

\subsection{Intermediate scales:  the transition regions}
\label{TransRegion_overview}

We refer to the regions between $\sim$10$-$40 kpc north and south of the 
galaxy as the North and South Transition Regions (NTR, STR). These 
regions are outside the main body of the galaxy,  beyond the
restarted Inner radio Lobes, extending approximately into 
the start of the Outer radio
Lobes. While the southern transition region is unremarkable,
the northern transition region hosts a collection of unusual
features, as follows.
 
 \subsubsection{Radio emission: the ``North Middle Lobe''}

 The North Transition Region is much brighter in the radio than the
 equivalent region to the south.  The radio-bright northern region,
 which we studied in Paper 1, is called the ``North Middle Lobe''
 (NML) in the literature.  In addition to the extended, diffuse radio
 emission, Morganti \etal (1999) identified a linear radio structure
 in the NML, and suggested that it might be a ``large-scale jet''
 connecting the inner and outer radio lobes.  However, in Paper 1 we
 showed that the linear feature does not appear to be a radio jet.  
 Rather, it is part of a knotty, radio-bright ridge embedded in the larger, 
 overpressured, diffuse structure of the NML.

The North Middle Lobe seems to have pushed aside the hot 
Interstellar Medium (ISM) of NGC 5128;  X-ray images 
reveal a cavity north of the galaxy, which is 
approximately coincident with 
the NML (\eg, Kraft \etal 2009;  Paper 1). 
We showed in Paper 1 that the radio-loud North Middle Lobe also 
needs ongoing energization: without a driver,
the NML will fade in $\ltw 20$ Myr due to expansion and
high-frequency synchrotron losses.

\subsubsection{``Weather'' in the north transition region}

In addition to unusual radio emission and the disturbed galactic
ISM,  the North Transition Region
hosts an X-ray-loud filament with embedded clouds, 
optical emission-line filaments, recently formed stars,
and a broken ring of cold atomic and molecular gas -- all in close
proximity.  As noted in Paper 1, we refer to this complex mixture of
radio emission, disturbed multiphase ISM, and star formation as the
{\it ``weather system''}.

\paragraph{Emission-line filaments and star formation}  
 A trail of emission line filaments, young star clusters, 
 and Ultraviolet emission (UV, $\lambda 1344-3000\AA$ )
 stretches outward from the galaxy.   Most previous optical 
 work has focused 
 on two bright regions close the galaxy, known as 
 the ``Inner'' and ``Outer'' Filaments.  The Inner Filament  
 is $\sim 8$ kpc from the core, 
 while the Outer Filament is $\sim 15$ kpc from the core.
 Both Filaments contain ionized gas, with high 
 internal/turbulent velocities ($\gtw 100$ 
 \kmps, ({\it e.g.}, Graham 1998), and so should notlast 
more than 10$-$15 Myr. 
 Both regions also contain young stars,
 with ages $\ltw 10$ Myr (e.g., Mould \etal 2000, Crockett \etal 2011),
 thus recent and ongoing star formation.
 In this paper, we show  
 that an extended ``weather ribbon'' of  
 UV and \hal emission extends much further, 
 reaching from the North Inner Lobe to at
 least $\sim 35$ kpc from the galaxy's core (see also Graham \& Price 1981).

\paragraph{Knotty radio/X-ray ridge}
The knotty, radio-bright ridge reported in Paper 1 sits close to
the SE edge of the North Middle Lobe.  It also sits very close to 
a similar, knotty, X-ray bright ridge (Feigelson \etal 1981,
Kraft \etal 2009).  We showed in paper 1 that the radio knots and X-ray
knots are spatially close to each other, but not exactly coincident; both
sets of knots are $\sim 1$ kpc in size and -- if they are diffuse
plasma -- are at $\gtw 20$ times higher pressure than their surroundings.
In this paper we show that the knotty
radio/X-ray ridge is also nearly coincident with the outer part of the
string of young stars and emission-line filaments mentioned above.

\paragraph{Cold gas and dust} The Transition Regions
 also contain clouds of dusty, cold gas  (\eg, Schiminovich 
\etal 1994), which form a broken ring orbiting NGC 5128. 
CO emission and warm dust 
have been detected from the densest parts of
the HI clouds (Charmandaris et al.  2000, Auld et al. 2012). 
One edge of the outermost HI cloud appears spatially 
coincident with the Outer Filament and with the radio 
and X-ray ridges.  Osterloo \& Morganti (2005) 
found that the cloud edge shows anomalous, turbulent 
velocities, and speculate that gas stripped from 
the cloud provides seed material 
for the emission line filaments NE of the cloud.

\subsection{Models for the transition regions}
\label{NML_models}

The astrophysics of the transition regions, as a whole, is 
not well understood, although various
models have been proposed for specific phenomena there.  

\paragraph{Star formation and emission line filaments}
There is near-consensus in the literature that an outflow 
must be causally related to the emission-line clouds and young stars 
in the weather ribbon NE of the galaxy.  Many authors assume  
the narrow feature seen by Morganti \etal (1999)
is a collimated jet which 
has energized the line-emitting clouds and induced star formation
in the ribbon.
Detailed models include  weak shocks in dense clouds 
photoionizing the ambient gas (Sutherland \etal 1993) and  shocks
causing cloud fragmentation and collapse leading to star formation 
(\eg, Fragile \etal 2004, Croft \etal 2006).   However, 
we showed in Paper 1 that the linear feature does not have the 
properties of a radio jet.  Although the lack of a detected 
jet seems to challenge  the jet-driven shock models, 
we argue in this paper that a broad wind outflow can play
the same role as a narrow jet in energizing the clouds and inducing
star formation.

\paragraph{Radio loudness of the North Middle Lobe}
We noted in Paper 1 that models proposed to 
explain the radio-loudness of the North Middle Lobe 
have either assumed the presence of a radio jet 
through the region, or attributed the NML to a slowly
rising, bouyant bubble.  We argued there  that none of these models
can explain all observed properties of the region -- in particular
 the short-lived nature of several
radio-related phenomena, the need for ongoing energy 
input to both Outer Lobes,  and the differences in 
radio properties between the North and South Transition
Regions. We suggest in this paper that a broad wind 
outflow can explain the radio properties of the North Middle Lobe, 
just as it can energize the emission-line clouds
and drive star formation in the weather ribbon. 


\begin{figure*}[htb]
{\center
\includegraphics[width=1.8\columnwidth]{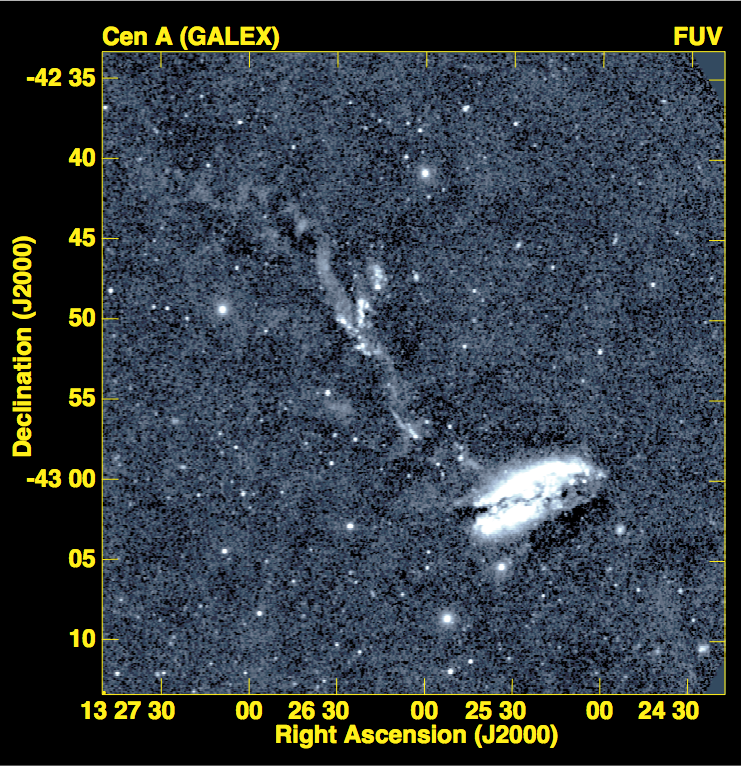}
\caption {FUV image  (from {\it GALEX:} $\lambda_{eff}= 1539 \AA$) 
of the region including NGC 5128 and the North
Transition Region.  The tilted central gas/dust disk is
clearly seen; FUV emission extends at least 1.5\arcmin above
and below the dust lane, and indicates a strong starburst 
in the disk.
The bright ``Inner'' and V-shaped ``Outer'' Filaments are also 
prominent in this image.   A fainter, twisted ribbon of FUV emission
extends out at least $\sim  35 \arcmin$ ($\sim 40$ kpc) 
from the center, and includes the Inner and Outer filaments. 
This image has been smoothed by a
$7.5\arcsec$ box to highlight diffuse emission.
}
\label{Fig:FUV_full_field}
}
\end{figure*}


\subsection{Organization of this paper}
\label{paper_org}

In this paper,  we present Galaxy Evolution Explorer ({\it GALEX}) 
observations of the inner 
$\sim 50$ kpc of Cen A/NGC 5128, and consider the likely astrophysics
of the Transition Regions.  In Section \ref{GALEX_Observations} 
we present new  {\it GALEX} observations which
reveal an extended ribbon of UV emission, extending well into the North
Transition Region, as well as 
 an active starburst in the central disk. 
In Section \ref{NML_WhatsTheStory}
we describe the complexities of the weather system, and show  
that nearly all of its features require ongoing energy input. 
In  Section \ref{Starburst_Power} 
we combine our FUV measurements with results from 
the literature to estimate the strength of the starburst in
the dusty disk. 
We argue that a wind from this starburst -- probably enhanced by
energy input from the AGN -- is 
flowing through the North and South Transition regions at present. 
In Section \ref{The_Wind_Section}
we explore a simple model of the wind. 
In Section \ref{Wind_Explains_Section} we argue that
the wind can drive the weather system 
and enhance the radio brightness of the northern transition region. 
Finally, in Section \ref{The_Last_Section} we summarize our results.


\smallskip

\section{{\it GALEX} Observations}
\label{GALEX_Observations}

UV  observations made by {\it GALEX} reveal a 
continuous UV-emitting ribbon extending 
more than 30\arcmin ~to the NE of the galaxy center,
and show that the dust disk of NGC 5128 is a strong UV source.

 {\it GALEX} is a NASA
Small Explorer mission optimized for wide-field 
ultraviolet imaging (Martin \etal 2005, Morrissey \etal 2005). 
{\it GALEX} images cover a $1.25^{\circ}$ diameter circular field,
with an angular resolution of $\sim 5\arcsec$.  
Far-UV (FUV, $\lambda 1344-1786\AA$, $\lambda_{eff} = 1539\AA$) and
Near-UV (NUV, $\lambda 1771-2831, \AA$,  $\lambda_{eff} = 2316\AA$)
images are obtained simultaneously, using a dichroic beamsplitter
and two photon-counting detectors.
Science observations are only made during the orbital nighttime.

Cen A was observed by {\it GALEX} for 14 eclipses between 
02-04 April 2004 and 01 May - 02 June 2008 in both FUV and NUV, 
and for a further 13 eclipses in NUV-only
between 02-09 April 2005.   Each exposure was processed using 
the standard {\it GALEX} pipeline (Morrissey \etal 2007) 
to reconstruct the image from photon lists, to un-dither the image, 
remove optical distortions and non-uniform detector response, to
remove internal reflections from nearby bright stars, and to
fit and subtract the sky background.  Individual background-subtracted 
images were then added to produce final combined images 
(total exposure time of 20072 sec in FUV, 30428 sec in NUV), 
which are available from MAST.\footnote{The Mikulski Archive for Space
  Telescopes; {\it GALEX} data available at galex.stsci.edu .}

    For this study, we extracted subimages including the 
galaxy and most of the North Middle Lobe detected in our radio 
observations (Paper 1).  The subimages were smoothed by a box 
$7.5\arcsec$ on a side, to enhance the faint UV emission, and 
for comparison with the much lower resolution radio images.  
In Figure \ref{Fig:FUV_full_field} 
we show the deep {\it GALEX} FUV continuum image
of a $40\arcmin$ (44 kpc) field around NGC 5128. The NUV image
is similar in appearance but is dominated by foreground stars,
making it more difficult to see the faint extended emission.

Figure \ref{Fig:FUV_full_field}  shows that 
a dramatic line of FUV emission extends
more than $30\arcmin$ ($\sim 35$ kpc) NE from the
galaxy center,  into the weather system in the  North Transition 
Region. This line of FUV emission appears to be a twisted curving ribbon;
it includes, but goes well beyond, the previously known
 Inner and Outer filaments.  We do not detect any corresponding extended
FUV emission SW of the galaxy.  (Diffuse NUV emission may
extend $\sim 15\arcmin$ ($\sim 17$ kpc)  SW of the galaxy 
(Auld et al. 2012), but it is right at the edge of the {\it GALEX} 
detector and is challenging to display.)


\begin{figure}[htb]
{\center
\includegraphics[width=.9\columnwidth]{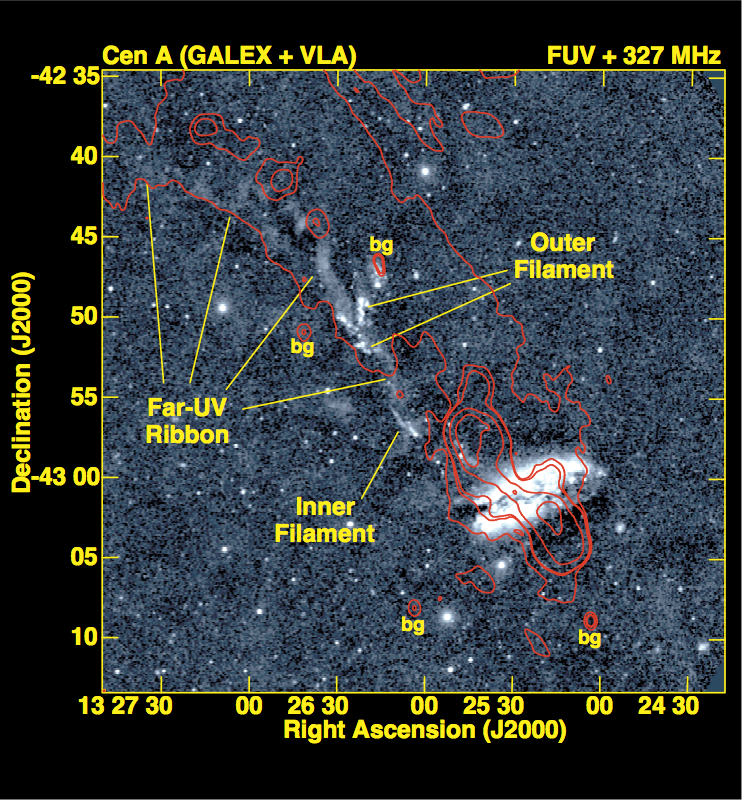}
\caption{ Overlay of 327MHz NML radio emission from Paper 1
(contours), superposed on the FUV {\it GALEX} image.   Radio
contours are at 0.07, 0.15, 0.18, 2.0, 10.0 Jy/beam. Key optical/FUV
features are labelled, as are four background radio sources (``bg'') 
unrelated to Cen A. The extended FUV ribbon overlays the radio ridge
running through the eastern edge of the diffuse radio emission,
including the radio knots.
Diffuse NML emission continues to the N and NW of this image, but
is not imaged well in these observations.  
The gap in the diffuse radio emission, approximately coincident 
with the Inner Filament, is real; the 327-MHz emission from this 
region is much fainter than from the rest of the NML.
}
\label{Fig:FUV_NMLradio}
}
\end{figure}


 In Figure \ref{Fig:FUV_NMLradio} we show the
 relationship of the UV emission to radio emission in the North
Transition Region.  We see that the twisted FUV ribbon sits 
close  to, but inside of, the SE edge of the diffuse 
radio emission;  it is also approximately coincident 
with the knotty radio-loud ridge (as indicated by the 
three radio knots, and as discussed 
in Section \ref{NML_Xray_radio_knots}).
The diffuse radio emission of the NML extends
 $\gtw 20\arcmin$ ($\sim 22$ kpc)
NW of the ridge as it connects to the North Outer Lobe;  
similar behavior is apparent at higher
frequencies (1.4 GHz, Morganti \etal 1999 and Feain \etal 2011;  
4.8 GHz, Junkes \etal 1993).  However, we see no sign of 
corresponding diffuse FUV emission in this region;
the FUV we detect is localized in the weather ribbon.

Figure \ref{Fig:FUV_full_field} also shows that 
the central disk of NGC5128 is a strong FUV source.
Because the FUV bandpass of {\it GALEX} is most sensitive to O and B stars,
it does an excellent job of identifying regions of recent star-formation.
This image reveals the bright (overexposed)
star-forming disk associated with the central starburst, 
along with the well-known dust lane cutting across the tilted disk.  
We see FUV emission above and below,
as well as along, the central dust lane. 
In {\it Hubble} Space Telescope images of the central $\sim 6$ pc of 
the disk (Marconi \etal 2000), individual young stars and clusters 
can be identified.
However, at the {\it GALEX} resolution ($\sim 5\arcsec$), we are
not able to tell how much of the FUV emission is direct 
starlight and how much is scattered.

\bigskip

\section{The Weather System: Probing the North Transition Region}
\label{NML_WhatsTheStory}

\begin{figure}[htb]
{\center
\includegraphics[width=.9\columnwidth, trim= 1cm 5cm 1cm 5cm clip=True] {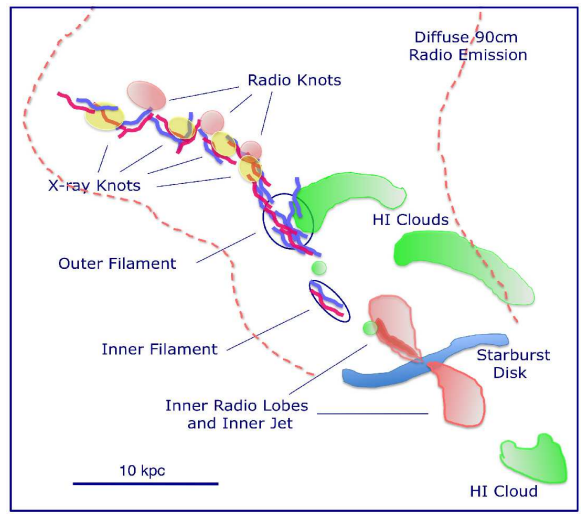}
\caption{Schematic diagram, showing the components of the ``weather system''
in the North Transition Regon of Cen A, and their apparent relationship
to the central galaxy. Each component is discussed in Section 
\ref {NML_WhatsTheStory}.  Blue and red wiggly lines indicate
FUV emission and H$\alpha$ filaments, respectively. }
\label{Fig:Cen_NTR_schematic}.
}
\end{figure}

The unusual, extended weather system north of NGC 5128
contains a complex mix of young stars, emission-line
clouds, diffuse radio emission and a knotty ridge of radio/X-ray 
emission. The collection of interesting oddities in the North 
Transition Region can be confusing.  We show the relationship of 
the various components
making up the weather system in Figure \ref{Fig:Cen_NTR_schematic}.
The various constituents of the weather system can provide 
important, if indirect, probes of the environment in the 
North Transition Region. 
As mentioned in Sections \ref{TransRegion_overview}
and \ref{NML_models}, none of the models so far proposed to 
explain these diverse phenomena works well for all of them.

In this section we discuss the properties and possible origins 
of the various features in the weather system.
We find that many of the interesting exotica are likely to have quite 
short lifetime, and to require very recent, probably {\it in-situ}
re-energizing.  We touch on how the weather phenomena might be 
related to a wind driven by the central starburst.  In following
sections we explore the likely galactic wind in more detail, and
show how a wind can drive many, if not all, of the observed weather 
features.


\begin{figure*}[htb]
{\center
\includegraphics[width=.85\columnwidth, trim=1.5cm 3.3cm 1.5cm 3.5cm, clip=True]{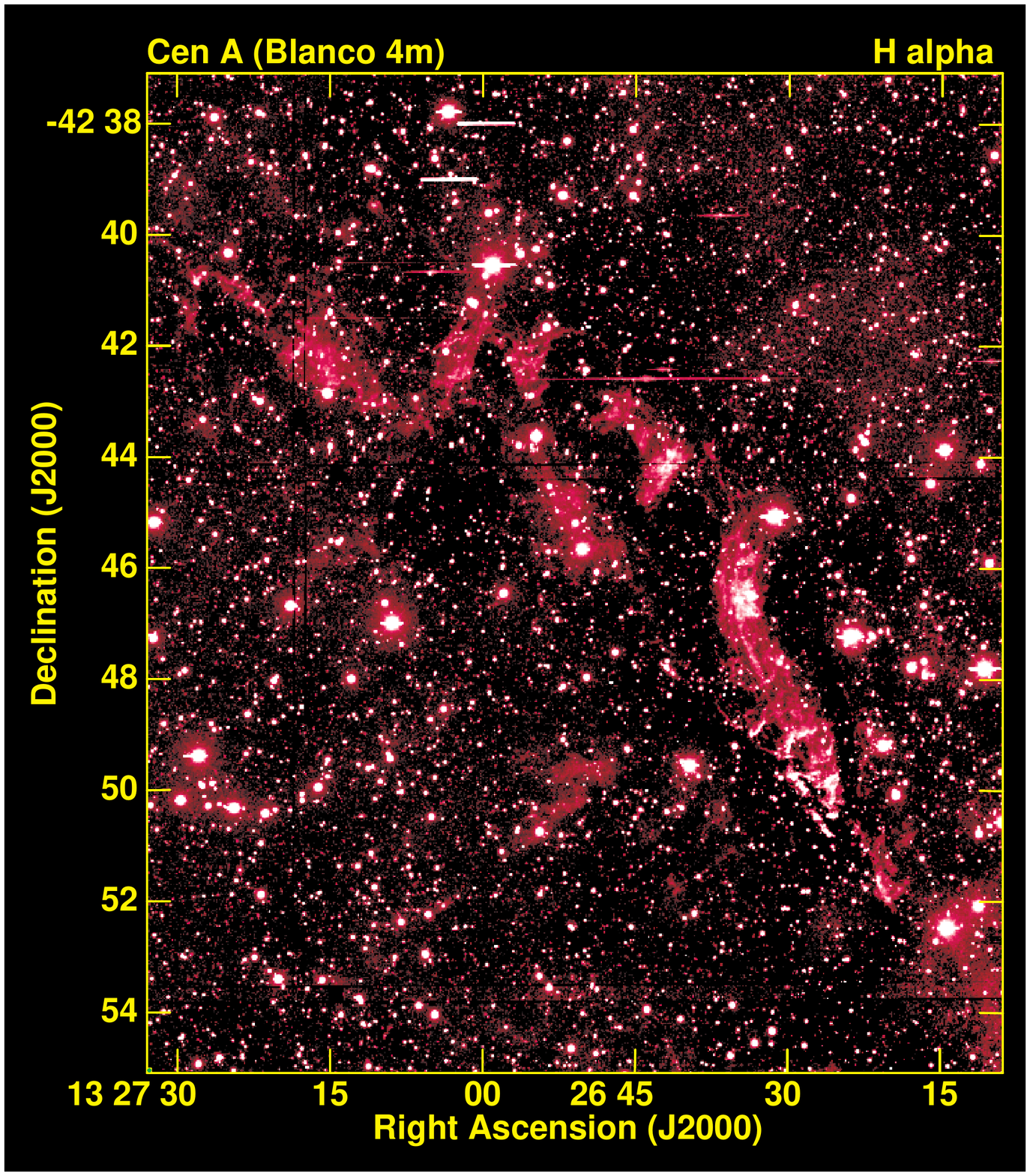}
\includegraphics[width=.85\columnwidth, trim=1.5cm 3.3cm 1.5cm 3.5cm, clip=True]{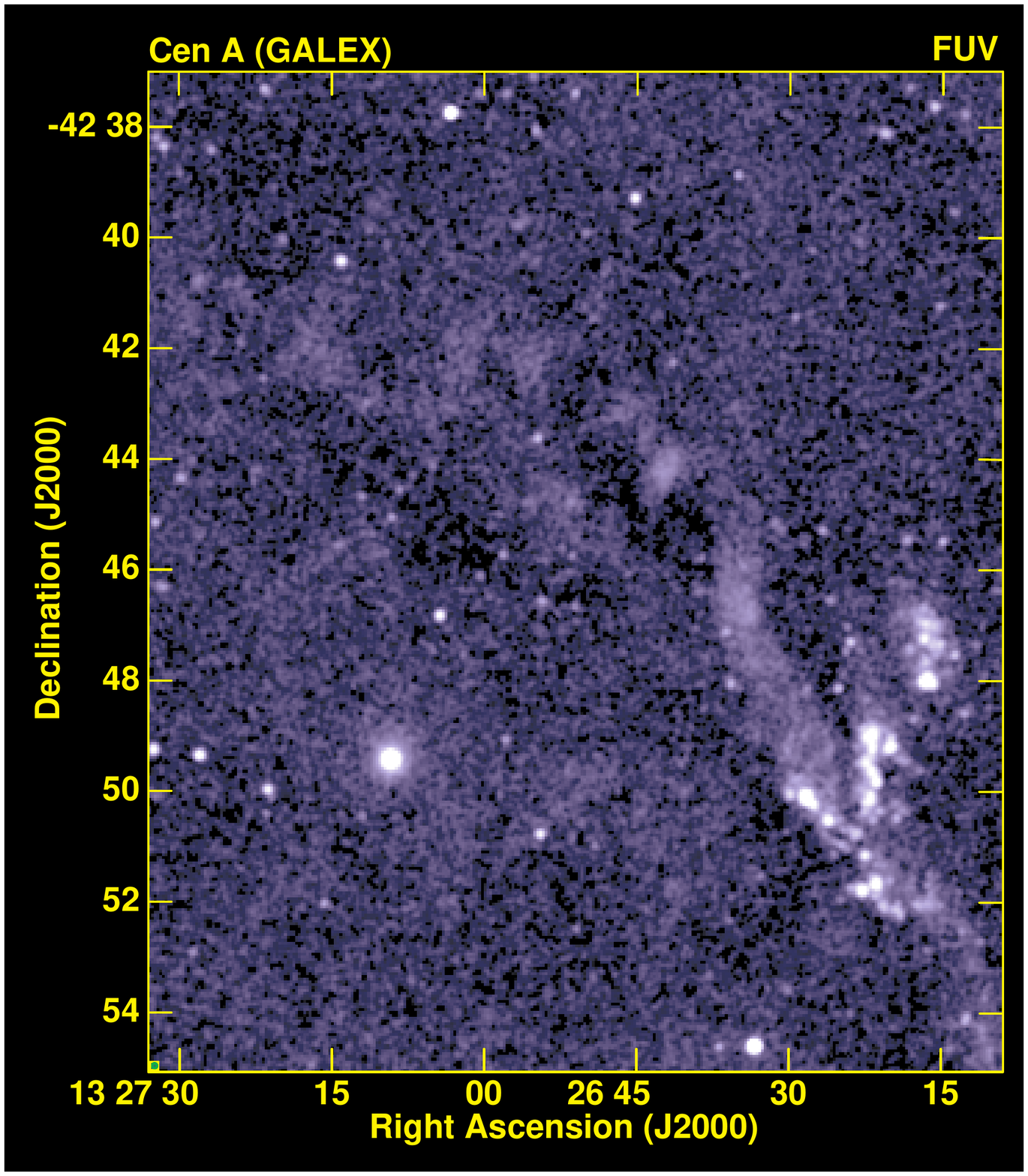}
\caption{The FUV weather ribbon, as seen in H$\alpha$ and FUV emission.
 Left: an  H$\alpha$ image of the weather ribbon, 
kindly provided by I. Evans and A. Koratkar.  A 1500 sec exposure, 
using a narrow-band interference filter centered at
$6009 \AA$, in $\sim 1\arcsec$ seeing.  
It has been smoothed with a $1.2\arcsec$ gaussian and then resampled 
with $1.5\arcsec$ pixels to match the FUV image. 
A background
gradient due to the host galaxy has been subtracted, using a 199-pixel
smoothing window 
($\sim 300 \arcsec$) to produce the subtracted
background.  
 Right, extracting the same field from
Figure \ref{Fig:FUV_full_field}, still with $7.5\arcsec$ smoothing to
bring out the faint emission.   Comparison shows that 
the diffuse FUV and H$\alpha$ ribbons are nearly identical, extending
from close to the galaxy (below and right of this image), through the V-shaped
Outer Filament (more easily seen in the FUV image, right),
and continuing in a twisted-ribbon structure
to $33 \arcmin (\sim 37$ kpc) from the galaxy center. 
}
\label{Fig:Optical_FUV_ribbon}
}
\end{figure*}


\smallskip

\subsection{Warm clouds and young stars}
\label{NML_filaments}

\paragraph{FUV and H$\alpha$ ribbon }
Both optical and FUV data reveal a complex mix of young stars, active
star formation and emission-line gas extending throughout the $\sim
30$ kpc eastern part of the NTR.  Figure \ref{Fig:Optical_FUV_ribbon}
shows a side-by-side comparison of H$\alpha$ and FUV images of the
weather system.  The left panel shows part of an H$\alpha$ image,
taken by I. Evans and A. Koratkar, using the MOSAIC2 camera on the 4m
Blanco telescope at CTIO.  The right panel of Figure
\ref{Fig:Optical_FUV_ribbon} shows the same field, extracted from our
full-field FUV image (Figure \ref{Fig:FUV_full_field}). The FUV and
\hal emission appear to be coincident, with the same extent and the
same morphology.

\begin{figure*}[htb]
{\center

\includegraphics[width=.85\columnwidth, trim=1.5cm 3.6cm 1.5cm 3.5cm, clip=True]{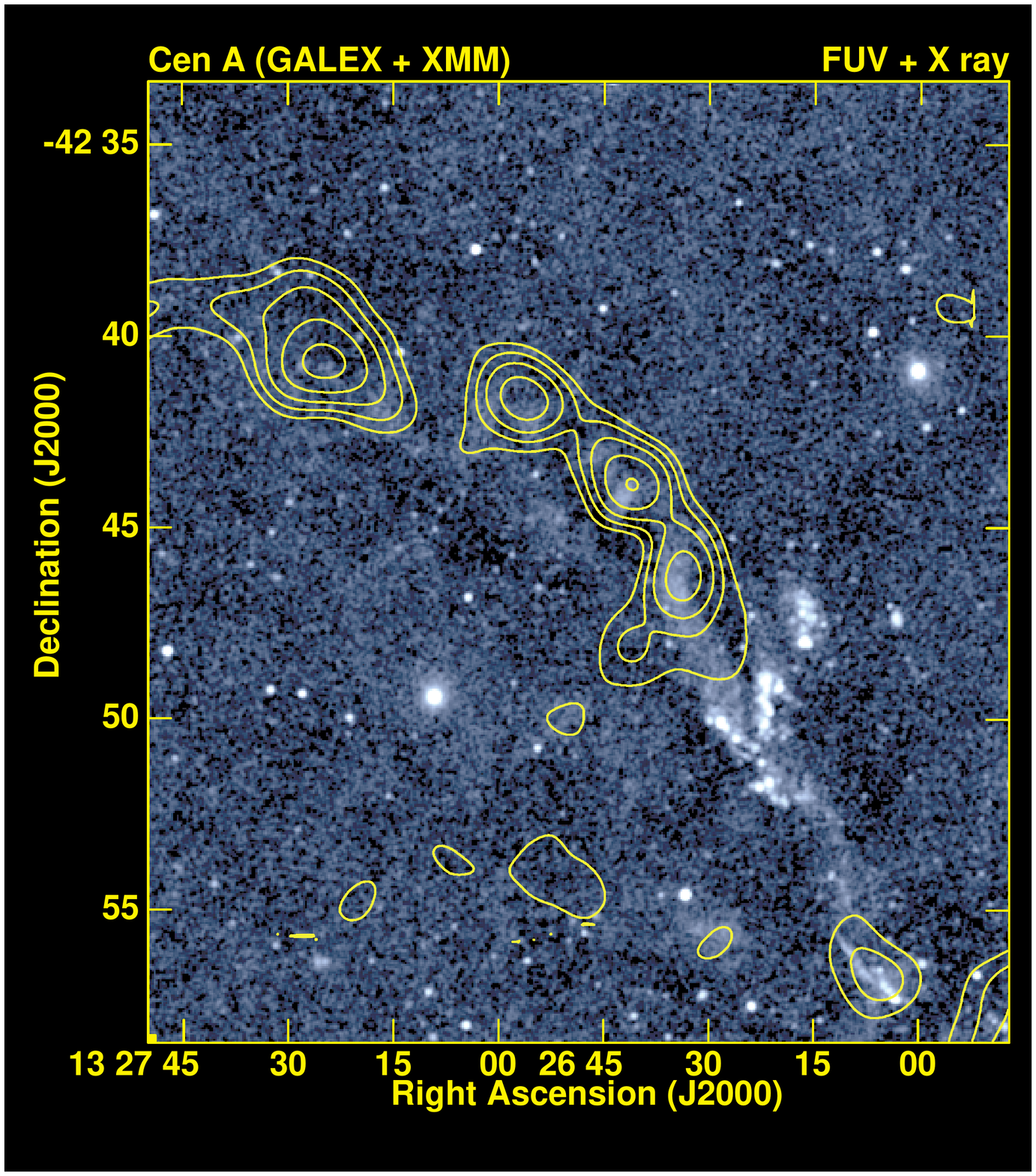}
\includegraphics[width=.85\columnwidth, trim=1.5cm 3.6cm 1.5cm 3.5cm, clip=True]{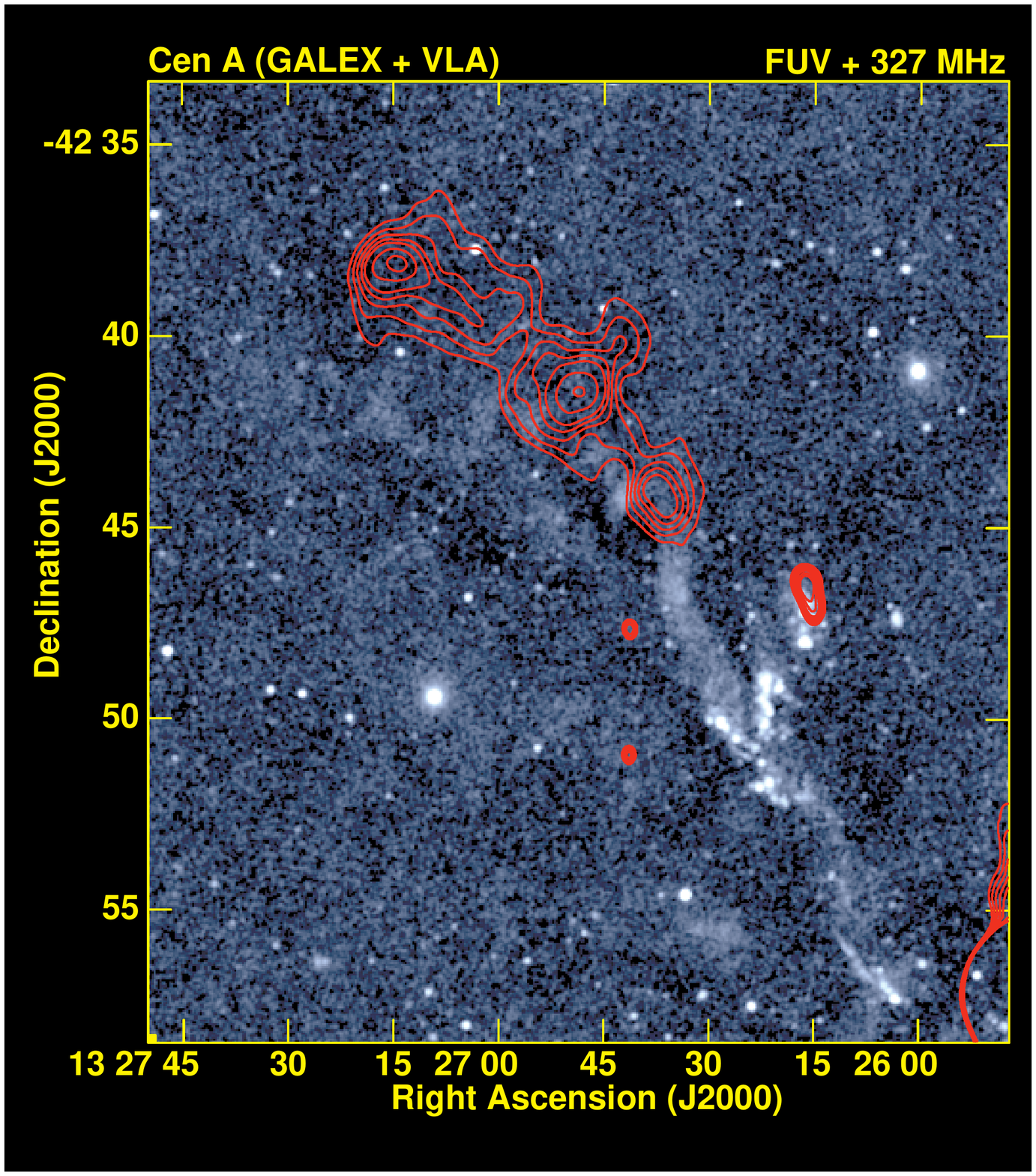}
\vskip 12pt
\caption{Left: Relationship between the X-ray knots (yellow contours;  from
K09;  smoothed XMM-Newton data kindly provided by R. Kraft)
 and the FUV ribbon (background image, our full-field {\it GALEX} image, 
from figure \ref{Fig:FUV_full_field}). The  
X-ray knots begin just north
of the V-shaped Outer Filament, and approximately track the
center of the twisting FUV ribbon.
Right: Relationship between the radio knots (red contours) and diffuse
FUV emission (background image). The three radio knots appear to be
associated with bends in the diffuse FUV ribbon (see also Figure
\ref{Fig:Optical_FUV_ribbon});  note that no compact radio emission is
associated with the Inner and Outer optical filaments.  (The three very 
compact sources S and W  of the radio knots are background sources).
Radio contours are from a 327 MHz image in Paper 1, with a beam size 
$28\arcsec \times 17\arcsec$.  Contour levels are at 130, 140, 150,
160, 170, 190, 210 mJy/beam. 
}
\label{Fig:Xray_FUV_radio_ribbon}
}
\end{figure*}

 The FUV/\hal
ribbon extends over $25\arcmin$ (28 kpc) 
starting near the galaxy
center and falling below our detection levels $\sim33\arcmin$ (37kpc)
from the galaxy center.  The comparison makes it clear that both
H$\alpha$ and FUV emission extend well beyond the bright emission line
filaments previously known (Inner and Outer Filaments, indicated in 
Figures \ref{Fig:FUV_NMLradio} and \ref{Fig:Cen_NTR_schematic}).  
The FUV/\hal ribbon contains the
well-studied Inner and Outer Filaments, as well as  \hal
emission regions farther away from the galaxy which were examined by
Graham \& Price (1981)\footnote{The V-shaped Outer Filament can 
be seen in the lower right of the FUV and \hal
images in Figure \ref{Fig:Optical_FUV_ribbon};   the Inner Filament is
closer to the galaxy and not included in Figure
\ref{Fig:Optical_FUV_ribbon} but can be seen in Figures
\ref{Fig:FUV_full_field}, \ref{Fig:FUV_NMLradio}, 
and \ref{Fig:Xray_FUV_radio_ribbon}.}.
FUV and H$\alpha$ emission, apparently part of the same large structure,
also extends inward from the fields shown in Figure
\ref{Fig:Optical_FUV_ribbon} towards the galaxy center  
(starting $\ltw 7\arcmin$ (8 kpc) from the galaxy center, 
as can be seen in Figure \ref{Fig:FUV_full_field}).

\paragraph{Kinematics}  The gas in the Inner and Outer Filaments,
 and in a few brighter regions in the outer ribbon, has a
 complex velocity structure (Graham \& Price 1981;  Morganti \etal
1991; Graham 1998).  The velocity of the  filaments in the line of sight
is $\sim 300-400$ km s$^{-1}$ away from the galaxy and towards us.  
In addition, turbulent internal velocities, of hundreds of \kmps,
are seen in the  filaments, as well as
gradients of $\gtw 200$ \kmps ~over only few hundred 
parsecs.  
Although no spectroscopy of fainter parts of the FUV/H$\alpha$ 
has been published, Sharp (2013, private communication) reports 
that the line-emitting gas just outside the Outer Filament 
is largely constrained 
to tight filaments in space and tight ranges in velocity.

We are not aware of any discussion in the literature on the origin of
these emission-lines' high velocities in NGC 5128.  However, we note
that clouds in several starburst galaxies (\eg, NGC 253, NGC 839, M82)
have similar velocities.  In those objects the clouds are thought to
be carried out by a starburst-driven hot wind.


\paragraph{ Likely Composition}
The diffuse, twisted ribbon appears nearly identical when
seen in H$\alpha$ and FUV; it is also visible in a very deep
{\it GALEX} Near-UV image (Auld \etal 2012).
It is probably a combination of warm line-emitting gas (\eg, Morganti
\etal 1991 or Sutherland \etal 1993) and very recent star formation (recent
enough that some of the newly formed stars still maintain HII regions,
\eg, Graham 1998, Mould \etal 2000, Crockett \etal 2012).  Our
current data do not allow us to discriminate between the two
possibilities, but it seems likely that both processes are involved.
 
The Inner and Outer Filaments are known to contain massive stars, some
of them very young ($\ltw 4$ Myr), and many of them exciting local HII
regions (Graham 1998, Mould et al. 2000,  Graham \& Fassett 2002,
Rejkuba et al. 2002, Crockett et al. 2011).  Our \hal and FUV images
of the outer ribbon show compact bright spots intermixed with the more
diffuse ribbon.  Some of these spots can be identified in both images;
this suggests they are HII regions around very young stars, embedded
in an extended warm ISM.  Some FUV-bright regions do not have H$\alpha$
counterparts (\eg, the extended west arm of the V-shaped Outer
Filament).

 FUV and H$\alpha$ emission may arise from diffuse clouds 
of warm gas; the close structural agreement of the FUV and 
H$\alpha$ ribbons suggests cospatial extended emission.  
Part of the ribbon may be composed of multiphase ISM 
structures, containing $10^4$K gas -- emitting \hal, [NII],
and [O III] lines -- and possibly 
$10^5$K gas, radiating in [CIV] at $1549 \AA$ 
(e.g., Werner \etal 2013, Sparks \etal, 2004).
High sensitivity Far-UV spectroscopic observations could
test this suggestion.

Some of the ionized gas cannot be excited by young stars alone.
The well-studied Inner and Outer Filament gas is 
characterized by a very high ionization level (Graham 
\& Price 1981, Morganti et al. 1991).\footnote{Sharp 
(2013, private communication) notes
that there is a noteable ionization gradient in part of the 
ribbon, beyond the Outer Filament, but that it will require 
detailed modelling to interpret.}
Young stars cannot account for this, 
because they reach at most  $\sim 15,000$K, 
while some of the ionized gas requires 
temperatures of $\sim 10^5$K.   

There is  disagreement in 
the literature about how the gas in the Inner and Outer
Filaments has been ionized.  Possibilities include photoionization from
the AGN (Morganti  et al. 1992), flow-induced shocks from a buoyant bubble
(Saxton et al. 2001), or shock-produced X-ray photoionization 
 driven by flow of a possible ``large-scale jet''
(Sutherland et al. 1993).
We note that many of the ionizing effects attributed to a nearby
jet could equally well be caused by a more diffuse fast flow, such as 
a galactic wind.

\subsection{Hot plasma:  Radio and X-ray Emission}  
\label{NML_Xray_radio_knots}

The optical/FUV ribbon is closely related to the knotty 
ridge, seen in radio and X-rays,
which sits near the SE edge of the North Middle Lobe. 
We showed in Paper 1  that the east side of the NML 
contains three distinct radio knots, each 1.3-1.5 kpc in size,  
connected by a radio-bright ridge which extends another 
$\sim10$ kpc inwards toward the galaxy from the knots.  
A similar, nearly coincident, structure is seen in soft X-rays 
(Kraft \etal, 2009). In paper 1 we noted three possibile
explanations for the radio and X-ray knots.  (1) They might be extant
clouds of cold gas which have been energized by local plasma
flow.  (2) They might come from interfaces between cooler emission-line
clouds and a hot surrounding medium.  (3)  They might be sites of
ongoing star formation;  their radio and X-ray power
suggests a rate $\sim 0.1 M_{\sun}$yr$^{-1}$.

In Figure \ref{Fig:Xray_FUV_radio_ribbon} we show a side-by-side comparison
of the X-ray and radio knots, both overlaid on the FUV field.
This makes it clear that the knot/bridge structures seen in radio and in 
X-rays are closely related to each other and also to the FUV/H$\alpha$ ribbon.

In the left panel of Figure \ref{Fig:Xray_FUV_radio_ribbon} we show that
the X-ray emission knots lie in close correspondence with the ribbon
of FUV emission. The X-ray ridge tends to track the center of the FUV
ribbon, and both systems show a similar southwards bend at the outer
end of the ribbon.  The X-ray knots coincide, within the limits of the
XMM resolution, with bright regions of the FUV/\hal ribbon.  This
correspondence of X-ray, FUV and H$\alpha$ emission also exists in the
Inner Filament, where Evans \& Koratkar (2004) detected X-rays, as shown
in the lower right corner
of the left panel of Figure \ref{Fig:Xray_FUV_radio_ribbon}.

In the right panel of Figure \ref{Fig:Xray_FUV_radio_ribbon}
 we compare the radio knots to the FUV
emission.  We find that the radio ridge does not correspond as 
closely to the ribbon as the X-ray ridge does.  
The radio ridge/knots are slightly offset to
the NW, and the three radio knots actually fall {\it in between} the
bright regions of the FUV/H$\alpha$ filament. This indicates that whatever
excites the radio knots is related to, but not spatially coincident
with, the warm ISM filaments seen in optical, FUV and soft X-rays.

Interestingly, Figure \ref{Fig:Xray_FUV_radio_ribbon} shows that 
both the radio and X-ray knots begin outside the V-shaped Outer Filament.
No soft X-ray emission is seen either from that structure, 
or from the FUV/H$\alpha$ ribbon between the Inner Filament 
and the Outer Filament.   The inner part of the radio ridge 
(galaxy-ward of the knots,  not shown here) sits close to, 
but not obviously associated with, the left arm of the "V" of 
the Outer Filament.
Although X-ray emission is associated with the Inner Filament,
radio emission is not.

\subsection{Cold gas: origin of the weather system?}
\label{cold_gas}

The origin of the warm 
gas in the weather system is not well understood.  The gas
may be initially cold gas which has somehow been injected into the
system, or it may be thermally unstable and cooling from the hotter
ambient ISM.  The first alternative is especially attractive for Cen
A, due to the presence of a broken ring of cold gas and dust clouds
which orbit the galaxy at radii of $\ltw 17$ kpc
(Schiminovich \etal 1994, Charmandaris \etal 2000, Auld \etal,
2012). Figure \ref{Fig:FUV_HI} shows the relationship between the FUV
ribbon and the HI clouds.  The cold material is thought to be left
over from a merger with a large gas-rich disk 250$-$750 Myr ago; the
same event provided the material that now forms the iconic central
dust lane ({\eg, Sparke \etal 1996, Quillen \etal 2006, Struve \etal
  2010).


\begin{figure}[htb]
{\center
\includegraphics[width=.9\columnwidth]{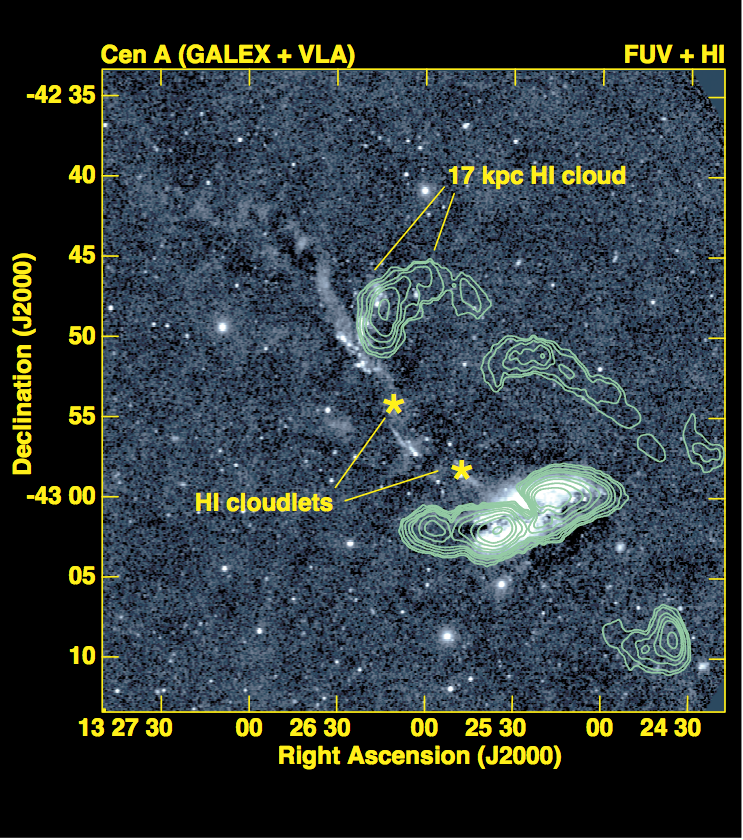}
\caption{ Overlay of HI contours superposed on our FUV {\it GALEX} image.
  HI data was obtained with the VLA in D array, image kindly provided
  by D. Shiminovich.  The HI velocities are consistent with a polar
  ring of cold gas, rotating with the central HI disk. The eastern
  edge of one of the northernmost HI cloud (labelled as
``the 17-kpc cloud'')  appears to abut the FUV ribbon,
  suggesting that it may provide some of the raw material forming the
  warmer ribbon. Two smaller HI clouds, discovered
by Struve \etal (2010) and marked here by asterisks, appear to
sit right on top of the inner line of FUV emission.}
\label{Fig:FUV_HI}
}
\end{figure}



\begin{table*}[htb]
\caption{ Timescales Relevant to the Weather System in Cen A / NGC 5128,
}
\label{Table:timescales}
\begin{center}
\begin{tabular}{l c c l}
\hline
\rule{0pt}{16pt} \hspace{-8pt} 
Region & Timescale & Reference & Comments
\\[4pt]
\hline
\rule{0pt}{16pt} \hspace{-12pt} 
Inner galaxy: & & &
\\[2pt]
dynamic age of inner lobes & $\sim(1-2)$ Myr &  \S \ref{Intro:large_scales}
& shock around SIL 
\\[2pt]
recent star formation  & 2 $-$ 50 Myr & \S \ref{Starburst_Power} & 
youngest stars in disk 
\\[2pt]
age of starburst &  $\gtw 50$ Myr & \S \ref{Starburst_Power} & 
oldest blue stars in disk
\\[2pt]
\hline
\rule{0pt}{16pt} \hspace{-12pt} 
Weather system & & & 
\\[2pt]
ages of young Stars &  4-15 Myr & \S \ref{Evanescence} & stellar 
colors/isochrones 
\\[2pt]
cloud radiative cooling time
 & $\sim$ 15 yr & \S \ref{Evanescence} & for $10^4$K emission-line cloud
\\[2pt]
emission-line cloud expansion time & $\ltw$0.6 Myr & \S \ref{Evanescence} 
& internal cloud pressure, unconfined
\\[2pt]
emission-line filament dispersal time & $\ltw$10$-$15 Myr & 
\S \ref{Evanescence} & cloud velocity dispersion 
\\[2pt]
radio/X-ray knot expansion time & $\sim 0.5\!-\!3$ Myr 
 & \S \ref{Evanescence} 
& if overpressure gas clouds
\\[2pt]
age of radio/X-ray knots & $\sim 10$ Myr 
 & \S \ref{Evanescence} 
& if star-forming regions
\\[2pt]
Kelvin-Helmholtz cloud shredding time & 1$-$2 Myr 
& \S \ref{cloud_destruction_simple} & from linear growth rate
\\[2pt]
cloud evaporation by thermal conduction & $\gtw$ 2 Myr 
& \S \ref{cloud_destruction_simple} & Spitzer conductivity
\\[2pt]
transit time to end of ribbon, for filaments & 50 $-$ 85 Myr 
& \S \ref{Evanescence} & from observed cloud velocities 
\\[2pt]
\hline
\rule{0pt}{16pt} \hspace{-12pt} 
North Middle Lobe: & & & 
\\[2pt]
wind transit time through NML & 10$-$30 Myr & 
\S \ref{wind_model}
& assumes  2000$-$6000 \kmps  wind 
\\[2pt]
diffuse NML  & $\ltw$ 16 Myr & \S \ref{Evanescence}
& overpressure expansion if unconfined
\\[2pt]
synchrotron aging   & $\ltw 20$ Myr &  \S \ref{Evanescence}
& at/above 90 GHz 
\\[2pt]
\hline
\rule{0pt}{16pt} \hspace{-12pt} 
Outer radio Lobes: & & & 
\\[2pt]
shortest radiative lifetime  & few Myr & \S \ref{Intro:large_scales}
 & for $\g$-ray-loud electrons
\\[2pt]
re-energization  & $\sim 30$ Myr & \S \ref{Intro:large_scales}
 & decay time for internal turbulence
\\[2pt]
dynamic age   & $\sim$ 1 Gyr &  \S \ref{Intro:large_scales}
& from dynamic  models
\\[2pt]
\hline
\rule{0pt}{16pt} \hspace{-12pt} 
Dynamics  & & & 
\\[2pt]
time since last major merger & 2$-$4 Gyr & \S \ref{Intro:begin} 
& intermediate age stars in halo 
\\[2pt]
time since last (minor) merger/ disk settling time & 250 $-$ 750 Myr & 
\S \ref{cold_gas} 
 & disk modelling of HI \& IR; ``blue arc'' 
\\[2pt]
rotation time for HI ring & $\ltw$ 370 Myr & \S \ref{cold_gas}
 & V$_{rot}\gtw$250 \kmps 
\\[2pt]
\hline
\end{tabular}
\end{center}
\end{table*}


One of the clouds making up the ring -- which we
call the ``17-kpc cloud'' (labelled in Figure \ref{Fig:FUV_HI}) --
is of particular interest.
Several authors (\eg, Graham 1998,  Morganti \etal 1999) 
have noted that the eastern edge of this cloud 
appears to be spatially coincident with the western arm of the V-shaped
Outer Filament, and have suggested that cold gas ablated from the
HI cloud is the source of that filament.  Osterloo \& Morganti (2005) showed 
that the east edge of this cloud -- where it is in apparent contact with
the H$\alpha$ and FUV ribbon -- has anomalous velocities 
which agree  well with velocities
of the turbulent warm gas observed in the adjacent Outer Filament.  
Furthermore, part of the FUV/\hal ribbon has a velocity 
gradient consistent with that of the HI ring (Sharp, 2013, private 
communication).  Both of these observations support the 
the idea that the cold cloud is a likely source of some or all of
the ``downstream'' ionized gas.

FUV and H$\alpha$ emission is seen closer to the galaxy center than 
the 17-kpc HI cloud and the rest of the detected HI system,
but no likely cold gas reservoir is apparent for these regions.  
What might be the source 
of that warm gas?   Struve \etal (2010) recently detected
two small HI cloudlets, both located along the line of ionized gas
that forms the inner part of the weather ribbon (indicated in
 Figure \ref{Fig:FUV_HI}).
One of the cloudlets sits near where the inner jet 
destabilizes and flares out into the North Inner Lobe ($\sim 5$ kpc 
from the nucleus), and the other sits just outside of the 
Inner Filament ($\sim 11$ kpc from the nucleus). 
The velocities of the two new cloudlets are 
somewhat less than the systemic velocity, but they
are not as blue-shifted as the emission line 
gas in the Inner Filament.
These cloudlets may be the last remnants of 
a previous extension to the HI ring, most of which has 
already been transformed into the inner parts of the 
FUV/H$\alpha$ ribbon through interaction with an outflow.

\subsection{Much of the Weather is Evanescent}
\label{Evanescence}

Like terrestrial weather, the most striking features of Cen A's 
weather system are short-lived. The outer parts of the ribbon 
cannot have survived the 
transit out from the  parental 17-kpc HI cloud, let alone 
from the inner galaxy, 
and they certainly cannot be remnants of a historical merger that 
happened a few hundred Myr ago.  These weather phenomena must have been 
recently stimulated, by some local driver close to where we see 
them today. In this section, we explore the range of short 
timescale constraints acting in the North Transition Region.  
Because the range of various timescales relevant 
to the history of the inner galaxy and/or the weather system 
is large, we collect those timescales for reference in 
Table \ref{Table:timescales}.  We will show in Section 
\ref{Wind_Explains_Section}  how these timescales can 
be accomodated in the presence of a galactic-scale wind.

\paragraph{Young Stars}
The well studied Inner and Outer Filaments contain very young 
stars (age $\sim 4\!-\!15$ Myr;  Mould \etal 2000; Rejkuba \etal 2002, 2011;
Crockett \etal 2011).  Some of these (\eg, those in the Inner Filament
 and the  eastern branch of the Outer Filament)
 have associated HII regions; others 
(western branch of the Outer Filament) do not.  
Stars hot enough to ionize HII 
regions have typical lifetimes $\ltw 10$ Myr.

The FUV emission places less stringent constraints
on stellar ages:  we 
expect to see FUV emission from massive O stars for 
$\sim$ 10Myr and from B stars for up to 100Myr. 
Although we have no direct information on the 
nature of the outer  FUV/H$\alpha$ ribbon, 
the same lifetime arguments will apply if the 
ribbon emission comes from recently formed stars.

\paragraph{Emission-line clouds:  existence}  
Individual emission-line clouds are
subject to several disruptive effects. To estimate the 
associated timescales, which depend on cloud parameters,  we note
that the relatively large  ($\gtw 100$ pc) emission-line 
``clouds'' studied in the Inner and Outer Filaments (\eg, Graham \&
Price 1981) are actually aggregations of much smaller clouds 
(Morganti \etal 1991, also the \hal
image in the left panel of Figure \ref{Fig:Optical_FUV_ribbon}). 
Following the analysis of Morganti \etal (1991),  who studied
individual clouds 
in the Inner Filament and some parts of the V-shaped Outer Filament,
we 
define a ``typical'' emission line 
cloud:  size $a_c \sim 10$ pc, density $n_c \sim 30$ \cm3 and 
temperature $T_c \sim 10^4$K. We assume these numbers are also 
representative of emission line clouds further out in the weather system.

These clouds cannot last long in their present form, for 
at least two reasons.

1) A typical emitting cloud requires ongoing 
energization to offset rapid cooling and recombination.  
The radiative cooling time is extremely short: estimating the
cooling function as 
$\Lambda(T_c) \sim 10^{-22}$ erg cm$^3$/s for $T_c \gtw 10^4$K,
 we see $t_{cool} = k T_c / n \Lambda(T_c) \sim 15$ yr.  
This is a short enough timescale for  changes to
to be observable, if one has sufficient angular resolution.  
For instance, a 10 pc cloud would subtend $\sim 0.5\arcsec$, 
so that secular changes in  
such clouds would be accessible to  current telescopes
over the time that the filaments have been studied. 

2) A typical emission line cloud is at much higher pressure 
than the surrounding ISM.
The cloud has $p_c = 2 n_c k T_c \sim 8 \times 10^{-11}$\dyncm2
 -- significantly higher
than the pressure of the local galactic ISM 
($8.5 \times 10^{-13}$ \dyncm2, Kraft \etal, 2009)
If these clouds are not confined, they will expand
at their internal sound speed ($\sim 15$ \kmps);  thus a 10-pc
cloud will dissipate in only $\sim 0.6$ Myr (see also Morganti \etal
1991).

\paragraph{Emission-line clouds:  kinematics}  
Emission-line clouds  within both the Inner and Outer Filaments  show
high random velocities ($\sim 150\!-\!400$ \kmps;
Graham and Price 1981, Morganti et al. 1991) within small 
regions ($\sim 100$ pc).
These structures will disperse in $\ltw 10-15$ Myr
unless they are somehow confined. Less is known about velocities in
the weather ribbon beyond the Outer Filament.  Graham and Price (1981) 
measured
velocities in a few of the brightest regions of the ribbon, and found
velocity dispersions similar to those in the Inner and Outer 
Filaments.

\paragraph{Radio and X-ray knots}
We showed in Paper 1  
that {\em if} the  radio and X-ray knots are diffuse plasma clouds,
they are strongly overpressured, $\gtw 20 p_{ISM}$.
Thus, they will dissipate quickly if they are not confined.
The X-ray knots will survive no longer than $\sim 3$ Myr (paper 1,
also Kraft \etal 2009), while the radio knots will last no more
than $\sim 0.5$ Myr (paper 1).

Alternatively, we noted in paper 1 that the
the X-ray and radio knots might come from star-forming 
regions. Hard X-rays, if observed, could be an 
indicator of short-lived, massive stars with lifetimes $\sim 10$ 
Myr.  Because the
region has not been observed in hard X-rays, we cannot definitely
say that the knots must be sites of recent star formation;  but evidence
for such young stars elsewhere in the weather ribbon (\eg, Mould \etal
2000, Crockett \etal 2012) suggests that the X-ray/radio knots 
could contain similarly young stars {\it if} they are star-forming 
regions.

\paragraph{Transit times from the inner galaxy} 
The survival timescales for filaments and knots are much 
shorter than the time it would take for
gas from either the 17-kpc HI cloud or from
the inner galaxy to have reached the weather system. 
The weather ribbon extends $\sim 35$ kpc from NGC 5128, and
$\sim 18$ kpc from the HI cloud (both distances measured in
the sky plane). The optical filaments are moving away 
from NGC 5128 and towards us at $\sim 350 $ \kmps
(line of sight velocity;  Section 5.1).
If the cloud speed in the sky plane is comparable, and
if the ribbon material originated in the 17-kpc HI cloud, 
it would take $\sim 50$ Myr for the line-emitting gas to travel 
from the 17-kpc HI cloud to the outer end of the ribbon.
Alternatively, it would take  $\gtw 85 $ Myr for 
the material in the weather ribbon to travel
from the center of the galaxy to its present location.

It follows that any young stars in the outer regions of the weather
ribbon must have formed there.  They are too young to have been formed
close to the galaxy and transported out.  Similarly, the emission-line clouds
and radio/X-ray knots are unlikely to have lived long enough to survive
the journey from the galaxy in their present form.  They too must be
energized {\it in situ} in the weather ribbon.

\paragraph{Other short-lived radio phenomena}
In Paper 1 we pointed out three lines of further evidence for recent
energy transport through, and dissipation within, the North Transition
Region.  (1) We found there that the North Middle Lobe is at a higher
pressure than the surrounding ISM of the galaxy; if were a static
structure, it would expand away in $\ltw 16$ Myr.  (2) We noted that
the detection of the North Middle Lobe at 90 GHz (by WMAP; Hardcastle
\etal 2009) requires the electrons radiating at those frequencies to
have been re-energized no more than $\sim 20$ Myr ago.  (3) Because
the Outer Lobes must have been re-energized no more than $\sim 30$ Myr
ago to keep them shining in $\g$-rays as well as radio (Eilek 2014),
the necessary energy must have flowed through both Transition Regions
at least this recently.

\bigskip


\section{A wind from the central galaxy}
\label{Starburst_Power} 

In the previous section we showed that the diverse phenomena
which comprise the weather system in the North Transition Region
require {\it in situ} driving, because they are too short-lived
to have been transported out from NGC 5128.  The AGN is a likely
energy source, but without a (detected) jet on 10-30 kpc scales it is unclear 
how the energy can make it from the galactic core to the
weather system. 

In this section we explore an alternative
 mechanism:  a diffuse galactic wind.
The {\em GALEX} image shown in Figure \ref{Fig:FUV_full_field},
also  Figure \ref{Fig:FUV_inner_gxy}, 
shows that an active starburst is currently
underway in the dusty disk of NGC 5128. 
With this starburst, in addition to the AGN, we have two potential
power sources in the inner galaxy.  We will argue that the starburst
-- probably enhanced by energy from the AGN -- is driving a 
wind which can provide the necessary energy transport into and
through the Transition Regions.

\begin{figure}[htb]
{\center
\includegraphics[width=.95\columnwidth, trim=1.3cm 4.5cm 1.3cm 4.5cm, clip=True]{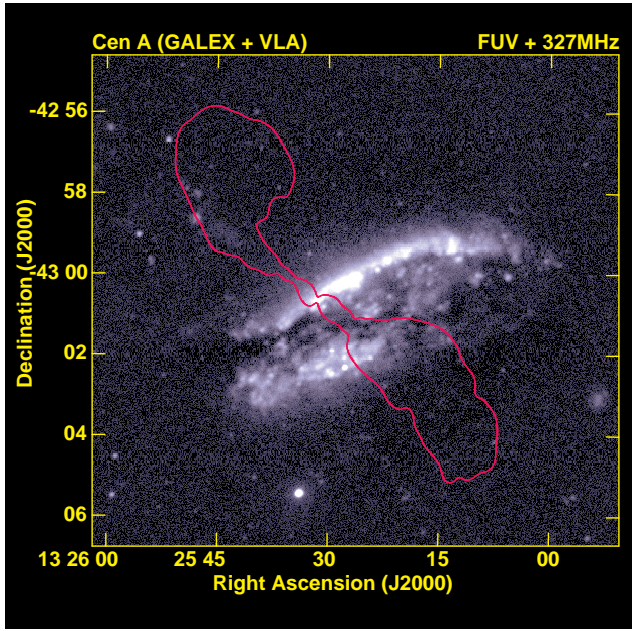}
\caption{FUV {\it GALEX} image of the inner 13 kpc of NGC 5128.  
The underlying image is an unsmoothed image, with resolution 
of $4.5\arcsec$. 
The red contour is the 1.1 Jy/beam brightness level in our unconvolved
90-cm image of the Inner Lobes (Paper 1). The  bright FUV
emission from the central gas/dust disk reveals a 
strong starburst in the central region of the NGC 5128. 
FUV emission is seen above and below the dust.
}
\label{Fig:FUV_inner_gxy}
}
\end{figure}


\subsection{Star formation rate in NGC5128}

FUV radiation reveals 
young stars not hidden by gas and dust clouds
(or their radiation scattered into our line of sight), 
while FIR radiation
traces young stars that are heavily obscured.  
We therefore combine FUV (1344-1786\AA),
FIR (8-1000$\mu$), and radio (90cm)  
measurements to determine the strength of the starburst.

\paragraph{Unobscured star formation: Far-UV}
We used the FUV image to characterize the unobscured 
part of the starburst. The burnt-out display in 
Figure \ref{Fig:FUV_full_field} shows that
FUV emission can be detected for $\sim 1.5\arcmin$ 
above and below the dust disk plane, and for a total 
distance of $\sim 8\arcmin$ along the dust plane.
The disk is quite thin, and warped multiple times 
(Quillen \etal 1993); 
Figure \ref{Fig:FUV_inner_gxy} shows the detailled 
structure of the central 13 kpc of the NGC 5128 
as seen in FUV. 

We used the AIPS program BLSUM to 
integrate the emission from the central galaxy 
in a hand-drawn region of size  $\sim 7.3\arcmin
\times 2.9\arcmin$ ($\sim 8.3 \times 3.3$ kpc),
enclosing all of the FUV emission shown in Figure 
\ref{Fig:FUV_inner_gxy} but excluding the 
base of the inner jet and nearby bright stars.  
We measured the local background in several nearly star-free
boxes near the galaxy (within 10$\arcmin$) and subtracted 
that level from the total measured flux.  Integrating across 
the {\it GALEX} bandpass, we derive the observed FUV flux
of the central starburst disk,
$F_{\rm FUV}^{\rm obs} \gtw 3.2 \times 10^{-12}$\ergps-cm$^2$/\AA.  
Multiplying  by the effective {\it GALEX} bandwidth (268\AA \, 
for FUV, Morrissey \etal 2007), and taking a 3.8-Mpc distance,
we estimate the directly observed FUV luminosity\footnote {Throughout 
this paper, L$_{\rm FUV}$ 
refers to the luminosity in the {\it GALEX} 
FUV bandpass,  $\lambda 1344-1786\AA$.} 
as $L_{\rm FUV}^{\rm obs} \sim 1.5 \times 10^{42}$\ergps.  

 To correct  for foreground Galactic extinction, we 
use $A_{\rm FUV} = 0.91$ mag (Gil de Paz \etal, 2007).  
Thus, the unobscured FUV luminosity from
the galaxy is boosted by a factor $2.3$, to $L_{\rm FUV}^{\rm corr}
\sim 3.5 \times 10^{42}$\ergps.  We convert this to a star formation
rate (SFR) with the proxy relation ${\rm SFR} 
 = 4.4 \times 10^{-44} L_{\rm FUV}$ 
(Murphy \etal 2011;  units of SFR are $M_{\sun}/{\rm yr}$
and units of $L_{\rm FUV}$ are \ergps in {\it GALEX} FUV bandpass). 
 We thus estimate the 
 {\it unobscured} star formation rate in the disk of NGC 5128: $({\rm
 SFR})_{naked} \sim .15 M_{\sun}$/yr.  If UV light escapes preferentially
above and below the disk, and is not scattered into our line
of sight, $({\rm SFR})_{naked}$ may be higher.


\paragraph{Obscured star formation: Far-IR}

The strength of an obscured starburst is commonly 
determined from its bolometric IR luminosity 
($L_{\rm FIR}: 8-1000 \mu$).  To estimate $L_{\rm FIR}$ 
for NGC5128, we started with its IRAS fluxes, $\sim 217$ Jy at
$60 \mu$, and $\sim 501$ Jy at $100 \mu$ (Rice \etal, 1988). 
We converted these to $40-120 \mu$ flux, using ${\rm FIR} = 1.26
\times 10^{-14} \left[ 2.58 S_{60 \mu} + S_{100 \mu}\right]$ 
(e.g, Helou \etal 1985).  We then multiplied (FIR) 
by the $\sim 1.5$ conversion factor typical for FIR-selected 
galaxies (Yun \etal 2001) finding that the bolometric 
flux $L_{\rm FIR} \sim 9.3 \times 10^9 L_{\sun}$/yr.

Both Kennicutt (1998) and Murphy \etal (2011) give proxy relations to
convert $L_{\rm FIR}$ to a star formation rate (SFR), with small
differences reflecting different assumptions detailed conditions
within the starburst. Kennicutt gives ${\rm SFR} \sim 1.7 \times
10^{-10} L_{\rm FIR}$; Murphy \etal give ${\rm SFR} \sim 1.47 \times
10^{-10} L_{\rm FIR}$; here, SFR is measured in $(M_{\sun}/{\rm yr})$
and $L_{\rm FIR}$ is measured in solar luminosities. Taking a
rough mean of these, we estimate the rate of {\em obscured} star formation
in NGC5128 as ${\rm (SFR)}_{obsc} \sim 1.6 M_{\sun}$yr$^{-1}$


\paragraph{Total star formation:  Radio}

We also used our 90-cm radio image (Paper 1) as an independent 
check of the star formation rate in the central gas/dust disk.  
Because radio luminosity does not suffer from extinction, it 
should give the total star formation rate (obscured plus 
unobscured). We pointed out in Paper 1 that we detect a "ruff" 
of radio emission, transverse to the Inner Lobes and approximately 
coincident with the gas/dust disk (compare, e.g., Figure 1 or 6 
of paper 1 to  Figure \ref{Fig:FUV_inner_gxy} of this paper).  
We used the flux of this ruff as a separate proxy for the star 
formation rate, as follows.  We measured the mean flux density 
of the NW and SE regions of the disk (outside of the bright 
central AGN) as 165 mJy/beam, which we estimate includes a 
background of 25 mJy/beam.  We estimated the total disk area 
as $5.3\arcmin \times 1.5\arcmin$;  thus the total radio flux 
from the disk $\sim$ (140 mJy/beam)$\times$(56 beams) 
$\simeq 7.84$ Jy at 327 MHz.   Because of the uncertainties 
in estimating the radio background and interpolating across 
the AGN, this flux estimate is probably only good to a factor 
$\sim 2$. 

We scaled this flux estimate to 1 GHz using the mean 
spectrum  of star-forming galaxies in this frequency range, 
$S(\nu) \propto \nu^{-0.55}$ (Marvil \etal 2014).  This gives 
a 1-GHz radio luminosity 
$L_{1~ {\rm GHz}} \sim 7 \times 10^{28}$ erg s$^{-1} $Hz$^{-1}$ 
for the star-forming disk in Cen A.  We then used the proxy 
relation from equation (16) of Murphy etal (2011), which 
combines thermal and non-thermal radio emission to estimate 
a total rate as ${{\rm SFR} =5.8 \times 10^{-29} L_{1 {\rm GHz}}}$. 
 In this relation, the units of SFR are $M_{\odot}$/yr, the 
units of ${L_{1 {\rm GHz}} }$ are erg s$^{-1} $Hz$^{-1}$;  we have 
assumed $T = 10^4K$ and $\nu = 1$ GHz.    From this we estimate 
the radio-derived SFR in the disk of Cen A as $\sim 4$ $M_{\odot}$/yr.   
Although this rate is formally larger than the net SFR we derive 
from FUV and FIR, $\sim 1.75 M_{\odot}$/yr, we do not take the 
difference seriously because of the large uncertainties in our 
measurement of the radio flux.  We thus argue that the radio 
data support the FUV+FIR data, showing that the central disk 
in NGC 5128 is currently forming stars at a rate $\sim 2 M_{\odot}$/yr.


\subsection{The central starburst and its wind}
\label{starburst_wind}

Combining the obscured and unobscured star formation rates,
and comparing those results to our radio-derived estimates
of the total star formation rate,
we estimate the central disk in NGC 5128 hosts a 
starburst with total SFR $\sim 2 M_{\sun}$yr$^{-1}$ -- 
comparable to that estimated for the Milky Way, but concentrated
in a region only a few kpc across.

To estimate the age of  starburst, we ideally need a good 
census of the young and middle-aged stars 
in the region.  However, the disk in
NGC5128 has not been the subject stellar population
studies, probably because it  is such a crowded, obscured region.
A few studies of  blue stars and accompanying H$\alpha$ emission along the
edges of the disk  (Dufour \etal  1979, Bland \etal  1987,
Minitti \etal 2004) have revealed individual blue stars or clusters 
with ages from $2\!-\!50$ Myr.  This places a lower
limit on the age of the starburst, and shows that it is still active
today.  To place an upper bound on the starburst age, 
we note that starbursts in other galaxies are typically estimated
to last up to several hundred Myr (\eg, Crnojevic \etal 2011,
McQuinn \etal ~2009, 2010). It therefore seems likely that the starburst
in NGC5128 has been going on for at least $\sim 50 \!-\! 100$ Myr.

 As with starbursts in other galaxies, we  expect  the starburst in the disk
 of NGC5128 to drive out a wind. To estimate the strength of the wind we use 
 the range of starburst models of Strickland \& Heckman (2009), for an
 intermediate age starburst (20-50 Myr),  scaled to $\ltw 2
 M_{\sun}$/yr.  From this we estimate the total wind power, 
 $P_w \sim 1 \times 10^{42}$\ergps, and total
 mass  flux $\dot M_w \sim  0.5 M_{\sun}$yr$^{-1}$
 (to north and south of the disk;
 following Strickland \& Heckman 2009).  Thus, a
 pure-starburst wind would contribute only a few percent 
 of the power estimated to come from the AGN (as summarized in
 Table \ref{table:Powers}, using estimates  from Paper 1).

 Can we be certain this starburst drives out a wind?
 There is a general consensus
 (\eg, Strickland 2009, Lehnert \& Heckman 1996) that a star 
 formation rate $\gtw 0.1 M_{\odot}$/yr-kpc$^2$ is sufficient to 
 overcome the inertia of cold ISM in the starburst region and 
 drive out a wind.  While we do not know the internal structure
 of the starburst in NGC 5128 -- it is obscured by the disk -- 
 infrared imaging (Parkin \etal 2012, using {\em Herschel};
 also Rice \etal 1988, from {\it IRAS})
 suggests it is concentrated in the inner $\sim 5$ kpc of the disk.  If 
 we assume the $2 M_{\sun}$yr$^{-1}$ 
 starburst is uniformly distributed over a disk 5 kpc in diameter, 
 we derive an areal star formation rate $ \sim 0.1 M_{\sun}$yr$^{-1}$kpc$^{-2}$.
 As this is just at the nominal (emperical)
 threshold for wind driving, we argue that the
 starburst in the core of NGC 5128 is capable of driving out a wind
 and is likely to do so, albeit perhaps not as strong 
 as those in better known systems (\eg, M82 or NGC 1569).


 \begin{table}[htb]
 \caption{Power estimates for Cen A }
 \label{table:Powers}
 \begin{center}
 \begin{tabular}{c c c  }
 \hline
 \rule{0pt}{16pt} \hspace{-8pt }
 Region / Epoch & Power$^a$ & Timescale \\
 &   (\ergps) & 
 \\[4pt]
 \hline
 \rule{0pt}{16pt} \hspace{-8pt} 
Jets, present-day$^b$ & 
 $\gtw 2  \times 10^{43}$ & today
 \\[2pt] 
Inner lobes, over lifetime & 
 $\sim (2-6) \times 10^{43}$ & $\sim (1-2)$ Myr
 \\[2pt] 
Starburst wind, recently$^c$  & $\sim 1 \times 10^{42}$ & $\gtw 50$ Myr 
\\[2pt]
Outer lobes, over lifetime
& $ \gtw (2 \!-\!5)  \times 10^{43}$ & 
  $\sim$ 1 Gyr 
  \\[2pt]
 \hline
 \end{tabular}
 \end{center}
 $^a$ The total power to both sides of the galaxy
assuming symmetric north and south outflows.  Radio-based
estimates taken from Neff \etal
(2014, Paper 1).\\
$^b$ Assuming jets uniformly filled with electron-positron plasma.  \\
$^c$ Assumes the wind is driven by a $\sim 2 M_{\sun}$yr$^{-1}$ 
starburst, and is not augmented
by the AGN (Section \ref{starburst_wind}, this paper).
\end{table}


\subsection{Does the AGN augment the wind?}
\label{AGN_plus_wind}

At the present epoch, Cen A contains both an ongoing
starburst and a recently recollimated (or perhaps restarted) AGN.  
In the previous section we described a stand-alone wind, \ie, 
one driven only by the starburst.   However, this is 
likely to be too simple.  We expect that relativistic outflows
from the AGN will interact with, and perhaps energize, the wind.
We do not know the recent history of the AGN (mostly active?  mostly
quiet?), but we can bound the 
problem by two possibilities.

\paragraph{A long silence from the AGN} 
As we discuss in Paper 1, the inner radio source 
is $\ltw 2$ Myr old (Croston \etal 2009).  
One possibility is that
 the AGN has been dormant for much
or all of the duration of the starburst.  This could be causal, 
perhaps because supernovae-driven turbulence within the starburst
disk  disrupts the accretion flow, and/or removes substantial amounts of
 material that would otherwise have accreted onto the central
black hole  (\eg, Wild \etal 2010, Davies \etal 2007).

To be specific, we assume the AGN in NGC5128 has been quiet for $\sim
50-100$ Myr (the likely duration of the central starburst, also
the delay that Davies \etal ~(2007) suggest between 
  start of a starburst and the restarted feeding of the AGN).
 If this is the case, any radio-loud plasma from previous active
episodes of the AGN
would long ago have reached and/or dispersed within the Outer Lobes.
For instance, if the disconnected Inner Lobe plasma kept a
coasting speed $\sim 3000$ \kmps (comparable to the deprojected
advance speed of the Inner Lobes; \cf $\!$ paper 1),
it would reach 150 kpc in 50 Myr - well beyond the Transition
Regions discussed here.
In this case, energy and mass flow within the Transition Regions 
will be limited to
that of the basic, starburst-driven wind described above.  

While we can't prove or disprove this picture, we find it unlikely
for several reasons.   It would require us to be 
 catching the AGN at a special time, only 1-2 Myr since its
restart after a long silence.   In addition, we argue below 
(in Sections \ref{EnergyBudgetClouds} and \ref{EnergyBudgetRadio})
that a wind driven only by the starburst does not have enough power
to support the emission  line clouds in the weather system, or the
radio emission from the North Middle Lobe. Finally, 
essentially every double radio galaxy known hosts a currently 
active,  radio-loud AGN.   Even the rare radio
galaxies  without detected jets connecting the AGN to the radio lobes 
 have detectable jets on kpc scales 
(\eg, Fornax A,  Fomalont \etal 1989; 3C310, van Breugel \&
Fomalont 1984). If it were common for the AGN supporting 
a large radio galaxy to go dormant for $\sim$ 10\% of the age of the source,
we would expect to see a similar fraction of the radio
galaxy population with no radio core -- which is not the case.

\paragraph{An  AGN coexisting with the starburst} An alternative possibility
is that the AGN, recently restarted, has only been quiet  for a 
time short compared to the duration of the starburst.  Perhaps the
AGN ``burbles along'',  dying down and then reactivating
 every few Myr.  This may happen, for instance, if most of
the  starburst is sufficiently spread out within the inner few kpc of
the star-forming disk 
that it has little effect on the accretion flow close to the AGN. 
As we discussed in paper 1, as the restarted jets propagate into the
wind, different evolution paths are possible when they grow to
tens of kpc scales.  They may continue to be identifiable as jets,
without much effect on the ambient wind flow.  In this case, we might
expect ``jet fragments'' to continue to coast within the ambient wind
when the AGN again becomes quiet.  Alternatively,
the jets may turn into  more diffuse, turbulent plumes, with the potential
to impact and energize the local wind flow.  In this case we might
expect the jet plasma to lose its identity as a jet, and to mix with 
the ambient wind. 

As an example, assume the AGN has been dormant
 for only $\sim 5$ Myr.  If this is the case, the jet plasma from 
the previous cycle would still be within the Transition Regions
(3000 \kmps $\times$ 5 Myr = 15 kpc; 3000 \kmps is the 
estimated advance speed of the Inner Lobes, Paper 1), and would 
still be radio-loud (estimating the 
synchrotron lifetime at several tens of Myr;  \eg, paper 1).
Because we don't see any such jet remnants within the Transition Regions
in Cen A, we speculate that jets from the previous cycle
must  have disrupted, dispersed
and mixed with the ambient wind plasma.  It follows that the
mass and energy from the from AGN may have
been deposited locally in the wind
-- perhaps quite close to the base of the wind -- and thus 
became part of the diffuse wind flow.

\subsection{Are there alternative drivers for the weather?}
\label{AGN_nowind}

Given that the observational evidence for a wind is only modest,
and that the power of an unaugmented wind is weak compared to 
the AGN power, one might ask if some other mechanism can drive 
weather system.

\paragraph{Photoionization}  
Morganti \etal (1991, 1992), suggested that the Inner and Outer
Filaments are photoionized by beamed radiation from a 
blazar.  However, Hardcastle et al. (2003) suggest that the
radio and X-ray brightness ratios in the inner jet appear 
inconsistent with a misdirected blazar model.  
Sutherland \etal (1993) point out that the ionization 
gradient noted by Morganti \etal (1992) does not 
point towards the nucleus, as might be 
expected if the Inner Filament were ionized by beamed 
EUV-Xray nuclear radiation.  Together with Viegas \& Prieto (1992),
Sutherland \etal (1993) also mention difficulties reproducing the 
observed [OIII]$I_{4363} / (I_{5007} + I_{4959})$ ratio 
using pure photoionization models.   
Crockett \etal (2012) note that a beamed photoionization model 
also fails to account for the highest excitation lines 
(\eg, Graham 1998, Evans and Koratkar 2004).
Furthermore, ionization by beamed radiation, by itself, 
does not explain the complex velocity field of the ionized 
gas in the Filaments. 

Relative orientations also do not support the idea
of beamed photoionization.  The inner jet is oriented at an an angle 
of  $30-41^{\circ}$ (depending on distance from the nucleus), 
with an opening angle of $\sim 8-10^{\circ}$. 
The Inner Filament is roughly aligned with the inner jet, with an
orientation of $\sim 28-33^{\circ}$ degrees, and thus it might might 
plausibly be ionized by beamed radiation from the nucleus or inner jet.
However, the jet is not aligned with the Outer Filament 
(orientation $\sim 44-53^{\circ}$), or with most of the outer 
ribbon structure (orientation $\sim 40-56^{\circ}$).  

Another possibility is that the FUV and emission-line regions are 
illuminated by the star-forming disk.  However, the agreement 
between radio-determined star formation rates and UV/IR-determined 
star formation rates (Section \ref{Starburst_Power}) 
precludes the possibility of significant amounts of hidden 
Far-UV emission being emitted along the dust-disk's axis.

\paragraph{Are filaments in cooling cores a similar phenomenon?}
Filaments of emission-line gas and young stars are also found 
close to central radio galaxies in cooling-core clusters 
(e.g., Fabian 2012;  Canning \etal 2014).  These filaments 
are similar to the weather system in Cen A:  they need 
ongoing energization, and they sit in a dynamic, hot 
($T \sim 10^7$K) ambient medium.  What can we learn from them?  
Although the filaments in cooling-core clusters sit in a denser  
ambient medium than that which surrounds the weather system in Cen A, 
there are interesting similarities.  The thermal state of 
filaments in cooling cores is far from understood, but there 
are several possibilities. They may be energized by energy 
transfer from the surrounding medium (e.g., heat conduction, 
Sparks \etal 2004,  or cosmic ray penetration, e.g. Fabian 2012).  
The filaments may also be energized dynamically, either by infall of the 
residual cooling flow, or by expansion work done by the radio 
galaxy if the filaments happen to sit close to the radio lobe
edges.

The cooling-core analogy is not perfect, however.  
The energetics of cooling cores are ultimately driven by a 
combination of slow cooling inflow and radio lobe expansion.  
Neither of these are clearly available for the Cen A weather system.  
There is no sign of cooling cores in small galaxy groups such 
as that which hosts NGC 5128.  The radio-loud plasma in both 
Transition Regions is more likely to expand, or flow, into the  
tenuous channels provided by the pre-existing Outer Lobes.  Thus, 
a different driver is needed for the weather system;  the wind 
we suggest is flowing through both Transition Regions can provide 
that driver.  If such a wind exists, the mechanisms by which energy 
is transferred to the cool gas clouds may be quite similar in 
Cen A and in cooling cores.  We discuss wind energization 
of the weather system in Cen A in more detail in 
section \ref{Wind_Explains_Section}.

\bigskip

\section{A wind through the transition regions}
\label{The_Wind_Section}

In the previous sections we have shown that many of the 
``weather''-related phenomena in the North Transition Region are 
short-lived, and require ongoing energization -- even though there seems
to be no AGN-driven jet in the region at present.  We have also suggested
that the starburst in the central gas disk, perhaps aided by power
from the AGN,  is driving out a strong galactic wind.  We speculate
that this wind can play the role of the (undetected) jet:  effects attributed
to jet-cloud interactions can equally well be caused by wind-cloud
interactions.  In this section we explore simple models of such a wind.
Although there is no robust detection of a hot wind in NGC 5128,
we can take advantage of what is known about 
the winds in some well-known starburst galaxies (\eg, M82,
NGC 253, NGC 2146) to understand the wind in NGC 5128.

\subsection{Envisioning the wind}
\label{Simple_wind_picture}

Basic  models of starburst winds (\eg, Chevalier \& Clegg 1985,
Strickland \& Heckman 2009) show that the wind
starts slowly, close to the starburst, but quickly accelerates and
becomes asymptotically supersonic. The key
property of the wind is that its terminal speed is a few times the
effective sound speed (governed by specific internal energy, both
thermal and nonthermal) in its core.  Detailed calculations for a
supernova-driven starburst wind  (\eg,
Strickland and Heckman, 2009) find $T_{ISM} \sim 10^8$K -- assuming
the starburst energy is efficiently thermalized and that there is not 
too much cold material in the nearby ISM.
For numerical estimates,
we assume such a wind has been going for $\sim 50-100$ Myr,
at power $P_w \sim 1.0 \times 10^{42}$\ergps, with mass
flux $\dot M_w \sim 0.5 M_{\sun}$yr$^{-1}$. 
  We conservatively\footnote{This assumes 
the wind is not totally ``cold'', but retains
$\sim 1/3$ of its power in advected internal  energy.}  
 estimate the asymptotic  wind velocity as $v_w \sim 2000$ \kmps.

If the starburst wind has been enhanced by relativistic
plasma from the AGN, we expect a hotter, faster wind. For a specific
example, let the AGN deposit $\sim 1 \times 10^{43}$ \ergps~ in the
plasma near the base of the wind (in the range of the likely total
AGN power, \cf ~Table \ref{table:Powers}).  Because the AGN outflow carries
very little mass, we assume the enhanced wind still has mass
flux $\dot M_w \sim 0.5 M_{\sun}$yr$^{-1}$, as estimated for the 
non-enhanced wind.   Because $ P_w \propto v^2$ at
constant $\dot M_w$, we estimate the wind speed as $v_w \sim 6300$ 
\kmps~ in the AGN-boosted wind.  

We note that these high wind speeds do not describe emission-line
clouds in the wind. Both observations (\eg, Forster-Schreiber \etal
2001) and modelling (Cooper \etal 2008, 2009) show the starbursting
disk is inhomogeneous, containing distributed star-forming clusters,
each of which drives out its own wind. The individual winds eventually
merge into a large-scale, diffuse starburst wind, which entrains and
accelerates streams and filaments of cooler, denser gas (from the
starburst disk or from interfaces with the local ISM).  The entrained
dense clouds do not attain the full speed of the hot wind, but they do
reach speeds of a few to several hundred \kmps, becoming the
emission-line filaments which characterize winds in well-studied
starburst galaxies (\eg, NGC 253, Westmoquette \etal 2011; NGC 839,
Rich \etal 2010).  The similarity of the cloud
outflow speeds in those galaxies to the speeds of the emission-line
clouds in the Cen A weather system (Section \ref{NML_filaments}) 
suggests that a similar wind may
blow through the North Transition Region in Cen A.

\subsection{Observational evidence for a wind}
\label{Obs_wind_evidence}

Is there any observational evidence of a galactic wind in NGC 5128?
Winds are easily identified in edge-on disk galaxies
by bipolar streams, extending several kpc above and below
the galaxy, detected in optical (emission-line
gas) and infrared (dust and cooler gas)  (\eg, 
 M82, NGC3079, NGC1482, NGC 253; Veilleux \etal 2005). 
Haloes associated with these winds can be detected 
at radio wavelengths, either as molecular lines from 
cold gas, (\eg, Sugai \etal 2003) or as synchrotron emission 
from a relativistic component,  (\eg, Seaquist \& Odegard 1999). 
Halos can also be observed in soft X-rays, 
from the diffuse wind fluid (\eg, Strickland and Heckman 
2009, M82), and inferred from detection
of [CIV] absorption at large impact parameters from starburst
galaxies (e.g., Borthakur \etal 2013). 

NGC 5128 does not display these canonical signs of a superwind.
However, it is not an isolated disk galaxy.  It is a large
elliptical, with a hot ISM, and it also hosts an AGN which has produced
the radio lobes of Cen A.  Thus, conditions in NGC5128 may be
sufficiently different from the well-known superwind host galaxies 
to preclude the detection of standard wind identifiers, or the 
wind signatures may be hidden by AGN phenomena.
For instance, except for the
singular weather system, NGC 5128 does not show the broad bipolar
outflow of warm and cold gas extending several kpc above and below the
galaxy.  This could be the result of less cold ISM being initially
available in the starburst disk, or due to increased stresses on
cool clouds as they propagate through the hot ISM of the elliptical galaxy.
Furthermore, detection of any radio or X-ray signature from the wind is
made difficult by several things:  
 the large angular size of Cen A (which makes both 
radio and X-ray observing challenging); the bright and complex 
radio and X-ray structures associated with the $\sim$ 5 kpc AGN jet; 
X-ray emission from the hot galactic ISM in NGC5128;  and 
interactions between the jet/Inner Lobes and the hot IGM.

Although the typical signatures of a superwind have not been 
detected in NGC 5128/Cen A, several observational results are 
suggestive of a wind:

1)  Philips \etal (1984) discovered a halo of highly ionized [OIII]
emission; they 
report this emission has velocity dispersion $\sim 350$ \kmps in the line of 
sight, and discuss its likely origin in a biconical outflow -- \ie, a wind --
from the central disk. 
The [OIII] is also detected by Bland-Hawthorn and Kedziora-Chudczer (2003).
Sharp and Bland-Hawthorn (2010) claim that the [OIII] closely matches
extended X-ray emission (reported by Kraft \etal 2008) extending  
$\gtw 1$ kpc above and below  the gas/dust disk, and $\sim 2$ kpc along disk.

2)  Emission-line imaging of clouds in the Outer Filament, 
and extending $\sim$ 8 kpc beyond the Outer Filament,
shows arcs of ionized gas which sit  ``upstream'' (galaxy-ward)
of the clouds.  
These structures are clearly visible in \hal and NII 
{\it Maryland Magellan Tunable Filter} 
images of the Outer Filament (Ellis, 2013, private 
communication\footnote{Images are visible online at
www.atnf.csiro.au/research/cena/documents/presentations/ellis.pdf}),
and can also be seen, albeit less clearly, in published images 
(Graham 1998 Figures 2 and 4;
Mould \etal 2000 Figure 4; Morganti \etal 1991 Figure 2b).  
The arcs are very suggestive of bow shocks expected to form upstream 
of dense clouds sitting in a supersonic flow.  These
images also show faint streamers of gas trailing NE of the 
ionized gas clouds, suggestive of material that has been 
removed from the clouds and is being carried along downstream 
by a wind.

3) Some features in X-ray images may reveal the base of the wind.
Feigelson \etal (1981) report two elongated X-ray ridges parallel to,
but offset from, the dust disk.  These are also apparent in 
recent Chandra images (Kraft \etal
2008 Figure 2, Croston \etal 2009 Figures 2 
and 3)\footnote{The most detailed image, from
Burke \etal (2013),  is available online at
www.chandra.harvard.edu/photo/2014/cena .}
which show a bright, diverging region
of X-ray emission, extending above and below the dust disk with
sharp edges suggestive of the walls of a superwind. Spectral
analysis of one of the walls (Region 3, Croston \etal 2009) finds
thermal emission from gas at $\ltw 1$ keV, not inconsistent with
the relatively cool edges inferred for other winds (\eg, Ranalli
\etal 2008).

\subsection{Does the wind reach the transition regions?}
\label{wind_model}

To learn how the wind propagates in and through the Transition Regions,
we use the two example winds from Section \ref{Simple_wind_picture}.
We choose a low-power, pure-starburst wind (at $P_w \sim 1 \times 10^{42}$ \ergps
~and $v_w \sim 2000$ \kmps),  and a higher-power, AGN-enhanced
wind (at $P_w \sim 1 \times 10^{43}$ \ergps
~and $v_w \sim 6300$ \kmps).  

Either  wind can easily escape the  galaxy.
Globular cluster kinematics show the gravitating
 mass of NGC 5128 $\sim 1 \times 10^{12} M_{\odot}$ out to $\sim 40$
 kpc (Woodley 2010), giving an escape speed $\sim 330$ \kmps on that
 scale.  A wind moving at $\gtw 2000$ \kmps ~is not seriously restricted by
 the galaxy's gravity.  If the wind flows freely at this speed, in 50
 Myr it will reach $\gtw 100$ kpc from its origin, thus will have no
trouble reaching 
 the Transition Regions in the likely life of the central starburst
($\gtw 50$ Myr;  Section \ref{starburst_wind}).

The wind does not flow into vacuum, however.  Its progress will be
impeded by thermal pressure of the ambient medium, $p_o$, which in
this case comes from the ISM of NGC 5128 and tenuous plasma in the
large-scale radio lobes.  When the ram pressure ($\rho_w v_w^2$) of
the expanding wind drops to the level of the ambient pressure the wind
can no longer expand freely.  Instead, it decelerates by means of a
termination shock which sits at the distance $R_{\rm shock}$, where the two pressures
balance.  Past $R_{\rm shock}$, the shocked wind plasma forms a hot
bubble, slowly expanding into the ambient medium. Immediately
past the shock, the wind plasma slows to subsonic flow at $\sim
(.3-.4)v_w$,
and decelerates
even further as it approaches the outer edge of the bubble 
(\eg, Castor \etal 1975).

To estimate the location of the termination shock -- and thus whether
the wind is likely still to be supersonic when it moves through the
Transition Regions -- we consider two illustrative geometries.  In one
case, the wind might expand spherically into an isotropic ISM. 
Alternatively -- and likely more relevant for Cen A -- the wind may be
channeled into the pre-evacuated Outer Radio Lobes.

\paragraph{Spherical expansion into ISM}

If there were no pre-existing radio lobe, the wind would expand spherically
 into the undisturbed galactic ISM at density $10^{-3}$\cm3 
and pressure $\sim 8.5 \times 10^{-13}$\dyncm2
 (revealed by X-ray data; Kraft \etal 2009). 
In this situation, the termination shock sits at a position given by
 $R_{\rm shock}^2  \simeq P_w / 4 \pi p_o$. 
In a pure-starburst wind the shock 
sits at only $R_{\rm shock} \sim 7$ kpc from the core;  in an AGN-enhanced
wind it sits at $R_{\rm shock} \sim 13$ kpc from the core.  
Thus, by the time the wind plasma reaches the Transition Regions, it 
carries the same amount of power, but becomes
a subsonic breeze rather than a supersonic wind.

\paragraph{Wind channeled into radio lobes}

The situation in Cen A is likely to be more complicated.  The low-density
radio lobes (previously evacuated by the AGN) may constrain the wind to
flow within a relatively small solid angle, $\Omega \ll 4 \pi$.  
If this is the case -- and if the wind flow remains laminar within
the channel --  the termination shock sits at a position given by
 $R_{\rm shock}^2  \simeq P_w / 8 \pi fp_o$, where $f = \Omega / 4 \pi$,
and we assume half of the total power goes to each side of the galaxy.
Based on radio images, we estimate the wind has expanded laterally to
a width of $\sim$  24 kpc
when it is 30 kpc from the galaxy in the North Transition Region. 
Thus,
the ``open channel'' occupies only a fraction $f  \sim
5\%$ of the full solid angle that would be available to a wind expanding
spherically into an isotropic medium.

To estimate pre-wind conditions in the
radio lobe, we follow Eilek (2014), who argued that the Outer Lobes on
large scales are in pressure balance with the local IGM 
(thus the pressure drops from $\sim 8.5 \times 10^{-13}$\dyncm2 close to the
galaxy, to $ \sim 3.2 \times 10^{-13}$\dyncm2 past $\sim 100$ kpc from the
galaxy).  In this scenario, the  termination shock of a pure-starburst
wind sits at $R_{\rm shock} \sim 30$ kpc from the core, nearly reaching the
outer end of the weather ribbon.  In an  AGN-boosted
wind, the shock sits at $R_{\rm shock} \sim 65$ kpc. Thus
if the galactic wind is channeled into the low-density radio lobes, it
may still be supersonic when it reaches the North and South Transition Regions.

\paragraph{Could we see the termination shock?}
Finally, one might ask if these termination shocks
could be detected observationally. 
 Detection in X-rays seems unlikely;  the density
jump in such tenuous plasma would be very faint, and X-ray
imaging would be severely challenged by the large angular size, 
and variations in Galactic foreground emission.  One might also
expect the shock to be radio-bright, due to particle acceleration and
magnetic field amplification local to the shock.  Eilek (2014) has in
fact suggested that the bright filaments within the outer lobes (seen in
ATCA image from Feain \etal 2011) could be internal shocks, driven by
transonic outflow within the lobes.  The complex nature
of the radio-bright filaments makes it difficult to identify any one
structure as ``the''  terminal wind shock;  but if the wind flow becomes
turbulent as it propagates into the Outer Lobes, it may be the origin
of the observed radio-bright filaments.

\bigskip

\section{Discussion: The Wind Can Drive the Weather}
\label{Wind_Explains_Section}

A wide range of phenomena in the North Transition Region require
ongoing energization:  emission-line filaments, young stars, radio 
and X-ray knots, and the radio-loud North Middle Lobe itself. In
addition, energy flow through the North and South Transition Regions
is needed to keep the Outer Lobes shining in radio and $\gamma$-rays.
Although there seem to be no collimated jets in either Transition
Region, we have argued that a 
wind from the starburst, perhaps augmented by the AGN, is flowing through
both regions right now.  In this section, we explore
how the wind in NGC 5128 can create or sustain these weather-related
phenomena.

\subsection{Connecting the wind to emission-line clouds}
\label{wind_to_weather}

Several authors have suggested that a jet flowing past dense, cold gas
clouds can energize the clouds, producing the weather phenomena
in Cen A. We argue here that a supersonic wind can 
have the same effect.  We follow previous authors who
have considered the behavior of cold dense clouds in a starburst wind,
and apply their work to the specific context of Cen A.

\paragraph{Induced star formation}

Although the detailed physics remains unclear,  it has often been
 suggested that plasma flow 
past cold clouds will induce star formation in the clouds
(\eg, Fragile \etal 2004, Gaibler \etal 2012).
The flow could, or course, be either a jet or a wind.
The ``typical emission-line cloud'' we use in Section 
\ref{Evanescence} -- to represent
the  emission line clouds studied by Morganti \etal (1991) -- is not
gravitionally unstable.  However, 
we can speculate that some of the clouds, when perturbed and/or
fragmented by the passing flow, will cool rapidly enough to become
unstable and collapse to form stars (\eg,
Mellema \etal 2002, Cooper \etal 2009).

\paragraph{Bow shocks around the clouds}

When a supersonic wind  encounters slower clouds, bow shocks will
form upstream of the clouds.  Some clear examples  are apparent
in simulations from Cooper \etal (2009).  \hal images of the Outer
Filament in Cen A strongly suggest that such bow shocks exist upstream
of the emission-line clouds there (Section \ref{Obs_wind_evidence}). 
This has several interesting consequences for the local weather.
(1)  The upstream ram pressure, moderated 
by the bow shock, will help confine the high-pressure emission-line clouds
and radio/X-ray knots (if the latter are diffuse gas).
(2)  The hot shocked wind plasma around the clouds will be an X-ray source
(\eg,  Marcolini \etal 2005, Cooper \etal 2009), consistent
with the observed spatial correlation of X-rays with some emission line
clouds.
(3)  The shocks are likely to
accelerate electrons to relativistic energies, and also enhance
the local magnetic field (as we discuss in Section \ref{Wind_to_radio}), 
consistent with the observed spatial correlation 
of nonthermal radio emission with the FUV/\hal ribbon.

\paragraph{Sources for the gas in the weather ribbon}

In Section \ref{cold_gas}, we suggested that cold HI close to NGC 5128
-- in particular the ``17-kpc cloud'' and smaller structures closer
to the galaxy (Figure \ref{Fig:FUV_HI}) -- might be the source of the
material in the extended weather ribbon. If this is the case, gas
clouds torn from the parental HI clouds must maintain their
identity for at least 50 Myr, the travel time from the 17-kpc cloud to
the end of the weather ribbon.  This may be difficult, because the
clouds are subject to  destructive forces as they are carried
outwards by a faster, hot galactic wind.  They can be be evaporated by
thermal conduction as they sit in the hot wind.  They can also be
shredded by the Kelvin-Helmholtz instability as the hot wind flows
past them.

To check  this, we  estimate lifetimes for our typical emission-line
clouds (Section \ref{Evanescence}) against both effects 
(details and references are given in the Appendix).   We find the 
answer is uncertain.  Simple estimates suggest the clouds are destroyed
very rapidly, in only a few Myr.  This seems to 
imply that HI closer to NGC 5128 cannot be the only source of material
in the weather ribbon.  However, recent simulations identify
mitigating effects which may slow down the destruction,
and retain the cloud's identity, for longer than the lifetimes predicted
by the simple arguments.

Thus, while it is possible that  known HI clouds can be the source
for the entire weather ribbon, the question is not settled.
  Other more local mechanisms may be necessary. Some of the gas
 in the extended weather system may have come from other cold 
clouds -- not yet detected, or perhaps now destroyed
 --  which happen to lie farther than
17 kpc from the galaxy and in the path of the wind. 
Another possibility is that star 
formation in the weather ribbon close to the galaxy  created
young star clusters which continue to move outwards at the speed of their 
parental gas clouds.  Winds from these young clusters, driven by normal
stellar evolution,  may supply new gas to the weather system which we now
detect as part of the FUV/\hal ribbon.

 \subsection{Energy budget for the clouds}
\label{EnergyBudgetClouds}

 There are many ways to energize cool clouds within a hot,
 supersonic wind.  The wind's ram pressure can drive shocks into a
 cloud, heating the cloud as the shocks dissipate. If star formation
 is induced within the cloud, young massive stars will photoionize
 their environs. Thermal conduction will heat the cloud.
Ambient shear flow, coupled to the cloud 
by the Kelvin-Helmholtz instability, will drive random motions and
 turbulence that heat the cloud when they dissipate.

Detailed modelling of these complex
processes is beyond the scope of this paper, but we can constrain
the energy budget.  Following Sutherland \etal (1993), we
consider  emission-line clouds in the Inner and Outer Filaments.  The
very short cooling time of these clouds (Table \ref{Table:timescales})
means they need ongoing energization, which must ultimately come
from the wind flow. Taking the  H$\beta$ power for a large cloud complex
(collection of small, 10-pc cloudlets;  Section \ref{Evanescence})
 in the Inner and Outer Filaments as $L_{H \beta}\sim 10^{37}$\ergps
(Morganti \etal 1991), and estimating the total line 
power $\sim (50\!-\!100)L_{H \beta}$ (Sutherland \etal 1993), we see 
the wind must supply $\sim (5-10) \times
 10^{38}$\ergps ~to each such cloud complex. 
We estimate the wind flow has $\sim 6$ kpc radius when it encounters
the Inner and Outer Filaments, and take  $\sim (200 ~ {\rm pc})^2$ as
the effective area of the cloud complex  seen by the incoming wind.

With these numbers, a pure-starburst wind running at
$  5  \times 10^{41}$\ergps
~(to each side of the galaxy) would convey $ \sim 2 \times 10^{38}$\ergps
to the cloud complex.  This is not 
enough to maintain the line luminosity, even if all of
the incoming wind energy is efficiently transferred to the line emission.
More energy is needed -- and it is available if the starburst wind is
enhanced by energy from the AGN.  
For instance, if the AGN boosts the wind power by a factor of ten,
the cloud complex would see $\sim 2 \times 10^{39}$\ergps -- enough to
keep the clouds stirred up and maintain their emission-line power.

\subsection{Connecting  the wind to the radio emission}
\label{Wind_to_radio}

The North and South Transition Regions are asymmetric.
 The North Transition Region contains the dramatic weather 
ribbon, seen in optical, UV and X-rays, but the South Transition 
Region contains no similar feature.  We have argued
that a bipolar wind from the core of the galaxy
creates the weather ribbon in the North Transition Region
 when it energizes cold gas that happens to lie in its path.
By contrast, the 
lack  of a similar weather system in the South Transition Region
must be due to to a lack of cold gas clouds in the wind's 
southern flight path.

\paragraph{Why is the North Middle Lobe radio bright?} 
The North Transition Region contains the diffuse,
radio-bright  North Middle Lobe as well as the radio-loud knotty 
ridge that coincides with
the optical/FUV/X-ray weather system. 
Synchrotron radio emission from  the region will be enhanced when
the interaction of the supersonic wind with the cold
gas  generates shocks (as the wind is forced to change direction
and/or decelerate) and MHD turbulence (easily generated by
any perturbation to an MHD flow).   Shocks can
quickly energize relativistic particles and amplify magnetic fields
(\eg, Drury 1983, Schure \etal 2012).
Alfven waves within MHD turbulence can also accelerate relativistic
particles (\eg, Lacombe 1977, Eilek 1979).

If this is the case, the radio-bright ridge and knots coincident with
the weather system are due to localized particle acceleration and
field amplification where the wind encounters
cool gas in the weather system.  The enhanced radio
surface brightness which continues to the northwest -- and defines the
larger North Middle Lobe -- could be due to advection of shock-accelerated
plasma within the wind and/or local particle acceleration due to MHD
turbulence in the wind.

\paragraph{Why no equivalent South Middle Lobe?}
No comparable radio structure has been
detected in the South Transition Region, which contains only weak,
diffuse radio emission  (perhaps
a factor $\sim 10$ fainter than in the NML; Paper 1). If the 
radio-brightness of the North Middle Lobe is due to to the interaction of the
galactic wind with cold gas clouds which happen to be in the region,
then the asymmetry of 
the North and South Transition regions must be due to the
lack of corresponding cold HI to the south of the galaxy. 
We therefore suggest that the radio-faint South Transiton Region
region shows the  basic, unperturbed wind flow.
That is, some relativistic, magnetized plasma
(from the starburst, possibly enhanced by the AGN) 
is advected out with the wind. It radiates faintly as it goes, but
its radio emission is not amplified by any interaction with cold
gas clouds in its path.

In contrast to the North Middle Lobe, which we have argued is at 
a higher pressure than the 
ambient galactic ISM (Paper 1, also Section \ref{Evanescence}), 
there is no evidence that the radio-loud
plasma in the South Transition Region (STR) is over-pressured.  
In Paper 1 we found that the {\it minimum} pressure in the 
radio-loud North Middle Lobe is a factor $\sim4$ larger than 
that of the ambient ISM.   Because in Paper 1 we could only 
put upper limits on the radio power from the STR,  we could 
not estimate its volume emissivity (which is required to 
calculate the minimum pressure;  e.g., Paper 1 Appendix).  
We can, however, use the lower-resolution image of Junkes 
etal (1993;  kindly provided by N. Junkes)  to estimate 
the mean surface brightness in the STR to be $\sim 10\%$ of 
that in a comparable area of the NTR.  Because the minimum 
pressure is approximately proportional to the square root 
of the surface brightness,  it is a factor $\sim 3$ 
lower, on average, in the STR than in the NTR.  We 
therefore argue that current observations are consistent 
with the South Transition Region being in pressure balance 
with the ambient ISM -- and thus with the South Transition 
Region not being energized by interaction 
with dense gas clouds in its path.

This argument requires a particular geometry for  the cold HI clouds shown
in Figure \ref{Fig:FUV_HI}.
If the gas in the northern weather ribbon is indeed being dragged out 
of the 17-kpc HI cloud by a wind (which we envision as filling a 
flaring cylindrical region above and below the starburst disk),
most of that cloud must be located outside the
wind; otherwise it would seed warm emission-line gas everywhere along 
its length.  Thus, the observed  arc of  HI clouds must be 
either in front of, or behind, the suspected wind flow.  Since only 
one edge of the 17-kpc cloud appears to be
impacted by the wind, that edge of 
the cloud must be rotating into the wind flow.
Figure \ref{Fig:FUV_HI} shows there is also an HI 
cloud SW  of the galaxy.  If this southwestern cloud
were located in the wind flow,
we would expect to detect a Southern Middle Lobe and an 
equivalent southern weather system. 
Taken together, these arguments suggest the southwestern HI cloud, 
and  most of the ring of HI clouds (\eg to North and Northwest of center),
are   well away from the wind flow.

\subsection{Energy budget for the radio emission}
\label{EnergyBudgetRadio}

Detailed  modelling of the complex shock/turbulence 
physics leading to enhanced synchrotron emission is beyond the scope of this
paper, but we can check the energetics.  
The radio-bright North Middle Lobe  has excess pressure 
$\gtw 4 \times 10^{-12}$\dyncm2  (Paper 1, assuming the baryon/lepton
ratio there is similar to that in our Galaxy). 
Taking the measured area of the North Middle Lobe as a cylinder with 
radius 12 kpc and length 30 kpc, the excess energy is $3 p V \gtw 5
\times 10^{57}$ erg.  
We have argued that this excess energy comes from the galactic wind
encountering cold gas in the North Transition Region
which happens to lie in the wind's path. 

If the galactic wind is driven only by the central starburst, 
it must have continued at its current power 
for $300$ Myr in order to provide the extra energy, or even 
longer if some of the wind energy continues past the Transition 
Regions to power the Outer Lobes.  While it is not impossible for
the central starburst to have continued so long,  the much shorter
 timescales  of North Middle Lobe phenomena ($\sim $ 15-30 Myr; Table 
\ref{Table:timescales}) suggest that higher energy input rates may
be required.  In addition, we find it unlikely
that the AGN would have remained ``off'' for such a long period (a
substantial fraction of the dynamic age of the Outer Lobes;
Section \ref{Intro:large_scales}, also discussion in Paper 1).

Alternatively, if the AGN has boosted the wind power by a
factor of ten (our example in Section \ref{AGN_plus_wind}),  that
wind provides enough energy to create the North Middle Lobe in only
30 Myr (or longer if some wind energy continues into the Outer Lobes). 
Because the starburst has lasted for at least 50 Myr, and probably 
longer, we see that an  AGN-augmented wind easily is able to energize
the North Middle Lobe, 
even if some of its energy continues on into the Outer Lobes.
This scenario fits well with the frequent re-energization of
shocks and/or turbulence in the Outer Lobes (Section \ref{Intro:large_scales})
needed to keep them  shining in radio and $\gamma$-rays.
That energy must have come from the inner galaxy, by way
of the North and South Transition Regions. Although there seem to be
no jets at present within the Transition Regions, a diffuse wind can play
the same role, carrying energy and mass from the galactic core
into the large-scale radio lobes.



\section{Summary}
\label{The_Last_Section}

In this paper we have presented new {\it GALEX} observations 
which change and enhance our picture of the circumgalactic
environment around NGC 5128 /  Cen A.

$\bullet$ {\em The North Transition Region is an interesting place.}
In addition to the radio-loud North Middle Lobe, deep FUV and \hal 
images show that a  complex weather ribbon -- containing a mix of 
young stars, emission-line clouds, bright X-ray knots and a knotty radio-loud
ridge -- extends $\gtw 35$ kpc into the North Transition Region.  Most
features in this weather system are evanescent, with typical lifetimes
only a few Myr.  Because the NTR is almost certainly older than this,
some {\it in situ} driver must be re-energizing the weather phenomena.

$\bullet$ {\em The inner galaxy contains two energy sources:  a 
starburst and an AGN.}  Our {\it GALEX} data reveal a 
starburst in the central dust/gas disk, with 
star formation rate  $\sim 2 M_{\sun}$yr$^{-1}$. It has
been going for at least $\sim 50$ Myr, and probably -- 
 based on other well-studied starbursts -- for more than
100 Myr.  It seems likely that the AGN has co-existed with 
the starburst for at least part of the latter's life, probably
adding its power to the overall energy budget of the system. 

$\bullet$ {\em A galactic 
wind is blowing through both Transition Regions 
right now.} Although there seem to be no collimated jets 
in either Transition Region (\ie beyond $\sim$ 7 kpc from the core), 
the  central starburst should be driving out a 
wind.  
Unless the AGN has been inactive  for the duration of
the starburst -- which we deem unlikely --  energy and mass
 driven out from the AGN has probably contributed to the wind.  The
resulting enhanced galactic wind will carry
mass and energy into and through the Transition Regions.  

$\bullet$ {\em The wind can drive the weather in the North Transition 
Regions.}  This galactic wind  can energize and stimulate star formation in
dense gas clouds it encounters in
the North Transition Region, thus both causing and maintaining the extended
weather system.   The interaction of the wind with dense gas clouds
 can also accelerate relativistic electrons, causing
 the radio-loudness of the North Middle Lobe. The lack of any 
comparable weather system or enhanced radio emission to the south does
not mean there is no flow in the South Transition Region, but rather
that no dense gas clouds happen to sit in the southern wind's flow path.

\bigskip

At some level, the difference between an AGN-augmented wind
and a wind-enhanced jet is largely a semantic one, probably depending on
the details of the interaction between the two. 
Further spectroscopic imaging observations, and perhaps polarization
imaging, particularly of the striking FUV and H$\alpha$ ribbon,
would be invaluable in confirming the existence of the starburst
wind, and would allow us to gain significant insight into
energy transport in this nearby active galaxy.
It may well be that such a combined outflow is a common property of
low-luminosity AGN, but we are not able to detect the combination
in more distant systems.  Future understanding of Cen A, the nearest
active galaxy, will require consideration of the influences of an
active starburst and wind on the system.

\begin{acknowledgements}

We are very grateful to I. Evans, R. Kraft,  D. Schiminovich, 
and N. Junkes for sharing 
their data with us.  We  appreciate discussions about ongoing work
with R. Sharp and S. Ellis.  Thanks to E. Greisen for friendly 
software support throughout this project, and to   
D. Thilker, S. Wykes, J. Marvil and J. Ott for stimulating 
conversations.  
SGN thanks the NRAO in Socorro, NM, for hospitality during major parts of
this work.

SGN is very grateful to Chris Martin and the {\it GALEX} team.
This work is based on observations made with the NASA Galaxy Evolution 
Explorer ({\it GALEX}), which was operated for NASA by the California 
Institute of Technology under NASA contract NAS5-98034.
In this work we have made extensive use of both NASA's Astrophysics Data 
System (ADS; hosted by the High Energy Astrophysics Division at the Harvard
Smithsonian Center for Astrophysics), and the NASA/IPAC Extragalactic 
Database (NED; operated by the Jet Propulsion Laboratory, California
Institute of Technology, under contract with NASA)

We appreciate thoughtful comments from the referee, which have helped us 
improve the paper.

\end{acknowledgements}

\begin{appendix}
\section{How long can the emission-line clouds last?}
\label{cloud_destruction}

Dense, cool clouds moving relative to a hot ambient medium are subject
to disruption by the 
Kelvin-Helmoltz (KH) instability as well as ablation by thermal conduction.
 If the structures
in the weather system have come from HI clouds close to NGC 5128, they
must be able to survive the trip in the face of this adversity.
In this Appendix we estimate these timescales and discuss possible
mitigating effects which might enhance the cloud's longevity.   For
numerical purposes we use our ``typical'' 
emission-line cloud (density $n_c \sim  10$\cm3, temperature $T_c \sim 10^4$K,
 size $a_c \sim 10$ pc;  Section \ref{Evanescence}).

\subsection{Simple calcuations suggest rapid destruction}
\label{cloud_destruction_simple}

Our typical line-emitting cloud will evaporate when
thermal conduction transfers heat from the ISM into the cloud core.
Because the Coulomb mean free path is short for our cloud, classical
heat flux applies.  Following Cowie \& McKee (1977), also Marcolini
\etal (2005), the evaporation time for a cloud of radius $a_c$ and
density $n_c$ can be written $t_{evap} \sim 2 n_c a_c^2 k_B / \kappa$,
where $k_B$ is the Boltzmann constant, 
and the thermal conductivity $\kappa$ contains the important
physics. If $\kappa$ is given by the Spitzer value (\eg, Cowie \&
McKee 1977), and the ambient ISM at $T \sim 10^7$K, we expect
the cloud to evaporate in only  $t_{evap} \sim 2$ Myr.  

If our cloud is in motion relative to the ambient medium,
it is also subject to the Kelvin-Helmholtz
instability, which will  heat and fragment the cloud,
and thus cause its dispersal (\eg, Klein \etal 1994). 
The timescale for this
instability to  disrupt a cloud of radius $a_c$, moving relative
to the ambient medium at some $\Delta v$, is 
$t_{KH} \sim a_c (n_c/n_w)^{1/2}/\Delta v$, where
 $n_w$ is the density of the local wind plasma
 (\eg, Fragile \etal 2004).  If our 
emission-line-cloud  sits in a pure-starburst wind (as in Section
\ref{Simple_wind_picture}), we estimate
$\Delta v \sim 2000$ \kmps and $n_w \sim 5 \times 10^{-5}$\cm3 (estimating
a 6-kpc flow radius for the wind close to the 17-kpc HI cloud).
Under these conditions, $t_{KH} \sim 2.2$ Myr. If the AGN boosts
the wind by a factor of ten (as in Section
\ref{Simple_wind_picture}), with $\Delta v \sim 6300$ \kmps, 
a similar calculation gives  $n_w \sim 2 \times 10^{-5}$\cm3, and
$t_{KH} \sim 1$ Myr.

\subsection{Possible extensions of  cloud life}
 
Both the Kelvin-Helmholtz and thermal ablation timescales are 
significantly  smaller than the
$\sim 50$ Myr cloud transit time from the 17-kpc HI cloud to the end
of the weather ribbon (Section
\ref{cold_gas}).  This appears to challenge the possibility that the
weather system material originated in the 17-kpc HI cloud, or from
other HI closer to the galaxy.  However, uncertainties in the
models may extend the effective lifetime of the cloud.

One important uncertainty is the actual value of
thermal conductivity in the North Transition Region.
Observations of galaxy clusters suggest 
 $\kappa$ may be significantly lower than 
the Spitzer value (\eg,  Churazov etal 2001, Narayan \& Medvedev
2001).  This is consistent
with many other lines of evidence supporting collisionless transport effects,
as well as thermal isolation of the clouds by magnetic fields in the cloud-ISM
interface.  Because the
relevant physics is far from understood, we do not attempt to estimate 
an alternative $t_{evap}$, but just note it may be significantly longer than
the classical estimate.

Another uncertainty is the evolution of a Kelvin-Helmholtz-unstable 
cloud.  Recent
work identifies effects which  may lengthen the cloud lifetime.
Marcolini \etal (2005) find the temperature gradient 
set up by thermal conduction will slow KH growth.  In addition, 
if a cloud can radiate fast enough to offset  shear-induced heating, 
Cooper \etal (2009) find that KH-stripped material will cool to $\sim 10^4$K
and accumulate in smaller cloudlets which are drawn into the 
downstream flow to form filaments of cool (emission-line) gas. In the
simulations these cloudlets can last several times longer than the
linear KH instability time.

In Cen A, the combination of these two effects {\em may} lead to the
cool cloudlets keeping their identity as emission-line clouds for long
enough to reach the end of the weather system.  However, because
current simulations neither extend to such long timescales, nor track
the eventual fate (such as fragmentation, evaporation, star formation)
of KH-stripped cloudlets, we cannot say with certainty that the 17-kpc
HI cloud is the source of the entire extended weather system.

\end{appendix}



\end{document}